\documentclass[a4paper,onecolumn,superscriptaddress,11pt,accepted=2019-09-27]{quantumarticle}
\pdfoutput=1

\usepackage[english]{babel}
\usepackage{amsmath}
\usepackage{hyperref}
\usepackage[numbers,sort&compress]{natbib}

\usepackage[numbers,sort&compress]{natbib}
\usepackage{soul}
\usepackage{tabularx, booktabs,ragged2e,hyperref}
\usepackage{bbold}
\usepackage[svgnames,table]{xcolor}
\usepackage[tableposition=above]{caption}
\usepackage{blindtext}
\usepackage{graphicx}
\usepackage{verbatim}
\usepackage{color}
\usepackage{amsmath}
\usepackage{microtype}
\usepackage{placeins}
\usepackage{bm}
\usepackage{bbm}
\usepackage{amsfonts}
\usepackage{amsbsy}
\usepackage{amssymb}
\usepackage[normalem]{ulem}
\useunder{\uline}{\ul}{}
\usepackage{adjustbox,lipsum}
\usepackage[table]{xcolor}
\usepackage{caption}
\usepackage{subcaption}
\usepackage{float}
\usepackage{array}
\usepackage{multirow}
\restylefloat{figure}
\usepackage{upgreek}

\DeclareMathOperator{\Tr}{Tr}
\newcommand{\R}{\mathbb{R}}

\newcolumntype{Y}{>{\centering\arraybackslash}X}
\newcolumntype{C}{>{\centering\arraybackslash}X} % <-- modified

\newcommand{\beq}{\begin{equation}}
\newcommand{\eeq}{\end{equation}}
\newcommand{\bqa}{\begin{eqnarray}}
\newcommand{\eqa}{\end{eqnarray}}
\newcommand{\bal}{\begin{equation}\begin{aligned}}
\newcommand{\eal}{\end{aligned}\end{equation}}

\newcommand{\specialcell}[2][c]{%
  \begin{tabular}[#1]{@{}c@{}}#2\end{tabular}}

\newcommand{\erf}[1]{Eq.~(\ref{#1})}

\newcommand{\erfs}[2]{Eqs.~(\ref{#1})--(\ref{#2})}
\newcommand{\erfa}[2]{Eqs.~(\ref{#1}) and (\ref{#2})}
\newcommand{\arf}[1]{{App.}~\ref{#1}} 
\newcommand{\srf}[1]{Sec.~\ref{#1}} 
\newcommand{\crf}[1]{{Ref.}~\cite{#1}} 
\newcommand{\trf}[1]{Table ~\ref{#1}}

\newcommand{\frf}[1]{Fig.~\ref{#1}}

\newcommand{\dg}{^\dagger}

\newcommand{\bra}[1]{\left\langle #1\right|}
\newcommand{\ket}[1]{\left|#1\right\rangle}
\newcommand{\braket}[2]{\left\langle #1|#2\right\rangle}
\newcommand{\ketbra}[2]{| #1\rangle\langle #2|}

\newcommand{\pref}[1]{(\ref{#1})}
\renewcommand{\eqref}[1]{Eq.~\pref{#1}}

\newcommand{\blk}{\color{black}}

\newcommand\stPW{\bgroup\markoverwith{\textcolor{red}{\rule[0.5ex]{2pt}{0.4pt}}}\ULon}

\newcommand{\sch}{Schr\"odinger}

\newcommand{\no}[1]{}

\begin{document}

\title{Open quantum systems are harder to track than open classical systems}
\date{October 4, 2019}

\author{Prahlad Warszawski}
\address{Centre of Excellence in Engineered Quantum Systems (Australian Research Council),\\
School of Physics, The University of Sydney, Sydney, New South Wales 2006, Australia}
\orcid{0000-0003-1766-4295}
\email{prahladw@gmail.com}

\author{Howard M. Wiseman}
\address{Centre for Quantum Computation and Communication Technology (Australian Research Council), \\
Centre for Quantum Dynamics, Griffith University, Brisbane, Queensland 4111, Australia}
\orcid{0000-0001-6815-854X}

%\pacs{?, ?}
%\keywords{?}

\begin{abstract}

For a Markovian (in the strongest sense) open quantum system it is possible, by continuously monitoring the environment, to perfectly track the system; that is, to know the stochastically evolving pure state of the system without altering the master equation. In general, even for a system with a finite Hilbert space dimension $D$, the pure state trajectory will explore an infinite number of points in Hilbert space, meaning that the dimension $K$ of the classical memory required for the tracking is infinite. However, Karasik and Wiseman [Phys. Rev. Lett., 106(2):020406, 2011] showed that tracking of a qubit ($D=2$) is always possible with a bit ($K=2$), and gave a heuristic argument implying that a finite $K$ should be sufficient for any $D$, although 
beyond $D=2$ it would be necessary to have $K>D$. 
%that for some Markovian $D$-dimensional open quantum systems, a $K$-state classical apparatus can keep track of the system state, for some finite $K\geq D$, even when the system state stays completely pure.  This is achieved by adaptively monitoring the environment in such a way that the system state is stochastically, and discontinuously, steered among $K$ pure states (which are collectively known as a physically realisable ensemble).  
Our paper is concerned with rigorously investigating the relationship between $D$ and $K_{\rm min}$, the smallest feasible $K$. 
We confirm the long-standing conjecture of Karasik and Wiseman that, for generic systems with $D>2$, $K_{\rm min}>D$, 
by a computational proof (via Hilbert Nullstellensatz certificates of infeasibility).  
%, we answer an open question, of long-standing interest [R. I. Karasik and H. M. Wiseman, Phys. Rev. A, {\bf 84}, 052120].  
That is, beyond $D=2$, %we prove that there are generic 
$D$-dimensional open quantum systems are provably harder to track than $D$-dimensional open classical systems. 
We stress that this result allows complete freedom in choice of monitoring scheme, 
including adaptive monitoring which is, in general, necessary to implement a physically realizable ensemble (as it is known) of just $K$ pure states.  
%It is a surprising fact that it is possible for a $K$-state classical apparatus to keep track of the state of a $D$-dimensional Markovian open quantum system, for some finite $K\geq D$.  This is achieved by adaptively monitoring the environment in such a way that the system state is stochastically, and discontinuously, steered among $K$ pure states (which are collectively known as a physically realisable ensemble).  Our paper is concerned with the relationship between $K$ and $D$.  In particular, we computationally prove (by obtaining Hilbert Nullstellensatz certificates of infeasibility) that there are generic systems for which $K=D$-state ensembles are impossible, for $D>2$.  In that sense, there are generic $D$-dimensional open quantum systems which are provably harder to track than $D$-dimensional open classical systems, even given complete freedom in choice of measurement scheme.
Moreover, we develop, and better justify, a new heuristic to guide our expectation of $K_{\rm min}$ as a function of $D$, taking into account the number $L$ of Lindblad operators as well as symmetries in the problem. The use of invariant subspace and Wigner symmetries (that we recently introduced elsewhere, [New J. Phys. https://doi.org/10.1088/1367-2630/ab14b2]) makes it tractable to conduct a numerical search, using the method of polynomial homotopy continuation, to find finite physically realizable ensembles in $D=3$.  The results of this search support our heuristic. We thus have confidence in the most interesting feature of our heuristic: in the absence of symmetries, $K_{\rm min} \sim D^2$, implying a quadratic gap between the classical and quantum tracking problems. 
%\pw{the use of the $\sim$ symbol means scaling doesn't it?  Even with some symmetries we've shown that it still scales that way.  So perhaps we should not limit the statement to the absence of symmetry?} 
Explicit adaptive monitoring schemes that realize the discovered finite ensembles are obtained numerically, thus facilitating future experimental investigations.
\blk

%According to this heuristic, the minimum ensemble size $\sim D^2$ for large $D$, implying a quadratic gap between the classical and quantum tracking problems.  The use of invariant subspace and Wigner symmetries (that we recently introduced elsewhere, [New J. Phys. https://doi.org/10.1088/1367-2630/ab14b2]) makes it tractable to conduct a numerical search, using the method of polynomial homotopy continuation, to find finite physically realizable ensembles in $D=3$.  The results of this search support our heuristic.  Explicit adaptive monitoring schemes that realize the discovered finite ensembles are obtained numerically. 

\end{abstract}

\maketitle
\section{Introduction}
\label{intro}

Tracking an open quantum system requires measuring the environment to which the system is coupled.  In this way, the experimentalist gains knowledge of the quantum trajectory~\cite{WisMil10} followed by the system of interest.  For the case of perfect detector efficiency, no system information is lost into the environment and the system trajectory maps the path of a pure quantum state.  It is of interest to ask, how much memory is required to track a pure state trajectory of open quantum system?  The answer is typically that an infinite memory is required, due to the fact that generic monitoring schemes will result in a continuous quantum state trajectory that occupies a non-zero dimensional manifold of pure states.  Remarkably, this is not always the case: it has been shown~\cite{wiseman2001inequivalence,karasik2011many,karasik2011tracking,daryanoosh2016stochastic,warwis2018a} that, via the implementation of especially chosen system-dependent adaptive measurement schemes, quantum trajectories of some systems can be constrained to a finite number, $K$, of pure quantum states.  This has profound consequences for the memory requirements of tracking an open quantum system, as a classical device with only $K$ states (a `finite state machine'~\cite{gill1962introduction}) is sufficient to follow the quantum evolution.  In this paper, we will investigate the minimum ensemble size, $K_{\rm min}$, 
%{Shouldn't this, and many occurrences below, be $K_{\rm min}$?}
%\pw{I found this difficult to answer!  I feel like often we are talking about ensembles in general rather than specifically the minimal sized one.  For example, when we refer to heuristic $K\geq (D-1)^2+1$ we often use the language `one can expect a PRE to exist when...'.  That is, because of the sentence structures we use I'd be happy to keep it as $K$.  Do you have another specific example where you strongly believe $K$ is incorrect?} 
that is achievable, given complete freedom of measurement scheme.  In particular, we compare and contrast $K_{\rm min}$ with the dimension, $D$, of the quantum system and so address some long-standing open questions of interest raised in Refs.~\cite{karasik2011many,karasik2011tracking}.  

\blk
%Our work builds on the existing literature 
%
%
%Quantum state ensembles that comprise the pure state quantum trajectories of interest 

%are examples of `physically realisable ensembles' (PREs)~\cite{wiseman2001inequivalence,karasik2011many,karasik2011tracking,
%daryanoosh2016stochastic,warwis2018a}, with the categorisation existing in order to distinguish them from non-PREs that may describe the correct average system evolution but that do not represent the set of states obtained in any possible measurement scheme.  
%
%
%
%PREs may be infinite in extent, but these would require infinite memory to track, so are not our concern here.
%
%
%
%The manifold of pure states followed by the quantum system is termed a `physically realisable ensemble'   

%{In this paper, we investigate the tracking of an open quantum system when the measurement scheme on the environment is chosen so as to imbue the quantum dynamics with classical characteristics: the quantum system is always in one of a finite number of pure states, and undergoes jumps between them.  This topic has been explored previously wiseman2001inequivalence,karasik2011many,karasik2011tracking,daryanoosh2016stochastic,warwis2018a under the title of `physically realizable ensembles' (PREs).  In the current work we study further the differences of these PREs compared with classical ensembles, in order to answer a number of open questions raised in Refs. karasik2011many,karasik2011tracking.}

To elaborate further, we begin by discussing the tracking of an open quantum system that is {\it classical} in the sense that the finite ensemble of quantum states that form the system trajectory are mutually orthogonal.  This will serve to benchmark our considerations of generic open quantum systems.  
A canonical `classical' example is Einstein's original model of stimulated and spontaneous jumps, which, in modern language, describes an open quantum system weakly coupled to a bath.  In equilibrium, Einstein's theory involves jumps between energy eigenstates (for example, Bohr's stationary atomic and molecular states~\cite{doi:10.1080/14786441308634955}) that could in principle be monitored, so that the system could be tracked just by specifying which of these states the system occupies.  Implicit in this early model is that the ensemble size is equal to the number of accessible energy eigenstates, $D$.
Clearly, a $D$-state finite, resettable, classical memory is capable of tracking the state of the (effectively) classical system.  Additionally, there is only a single way to monitor the environment that provides complete information as to the transitions (for the atomic model this would be the photon number basis).

In contrast to the effectively classical case just discussed, the choice of monitoring of a generic open quantum system can have a profound effect upon its evolution.  The system, by definition, is interacting with the environment and, for suitable initial conditions, becomes entangled with it.  The measurement of the environment by an experimentalist effects `quantum steering'~\cite{schrodinger1935discussion} upon the system.  In this paper we are concerned with the case of continuous Markovian dynamics induced by the bath, also known as quantum white noise (QWN) coupling; the system will then, {\it in the absence of measurement}, obey a Lindblad-form master equation (ME) for the density matrix\blk~\cite{WisMil10}: 
\beq \label{me1}
\dot{\rho} = {\cal L}\rho \equiv -i[   \hat{H}_{\rm eff}\rho   -  \rho \hat{H}_{\rm eff}\dg] 
+ \sum_{l=1}^L  \hat{c}_l \rho  \hat{c}\dg_l,
\eeq
where $ \hat{H}_{\rm eff} \equiv \hat{H} - i\sum_l   \hat{c}\dg_l  \hat{c}_l/2$ and $\hat{H}$ is the Hermitian Hamiltonian. The purpose of separating \erf{me1} into terms that involve $ \hat{H}_{\rm eff}$ and those comprising $\hat{c}_l \rho  \hat{c}\dg_l$ is that they can be associated with a purity-preserving unraveling of the ME, as would arise from perfectly efficient monitoring of the decoherence channels (that are indexed by $l$).  Assuming an initially pure system state,  if a detection in channel $l$ is observed at time $t$ then the system jumps from the pre-jump state $\ket{\psi(t^{-})}$ to the post-jump state $\ket{\psi(t)}\propto \hat{c}_l \ket{\psi(t^{-})}$.  After the jump, the quantum state evolves under the no-jump evolution operator $ \hat{H}_{\rm eff}$ and will typically {\it not} remain stationary unless it happens to be an eigenstate of $ \hat{H}_{\rm eff}$.  

The stochastic path followed by the state is known as a quantum trajectory, and different monitoring schemes will lead to different types of quantum trajectories~\cite{CarQTraj,WisMil10}.  In fact, there are an infinite number of ways to measure the environment that maintains a pure quantum state for the system.  This follows from the invariance of \erf{me1} under the following joint transformations~\cite{PhysRevA.49.2133,karasik2011many}
\bqa
 \hat{c}_{l}&\rightarrow&\left\{ \hat{c}^{\prime}_{m}  = \sum_{l=1}^{L} S_{ml} \hat{c}_l + \beta_m\right\}  \label{jumpOps}\\
\hat{H}&\rightarrow&\left\{ \hat{H}^{\prime}= \hat{H} - \frac{i}{2}\sum_{m=1}^{M}  (\beta^{*}_m   \hat{c}^{\prime}_m - \beta_m  \hat{c}^{\prime\dagger}_m{}) \right\},
 \label{noJumpOps}
\eqa
where, with $M\geq L$, $\vec{\beta}$ is an arbitrary complex $M$-vector and ${\bm S}$ is an arbitrary $M\times L$ semi-unitary  matrix; that is, $\sum_{m=1}^{M} S^{*}_{ml}S_{ml'} = \delta_{l,l'}$. 
By unraveling \erf{me1} with $\{\hat{c}^{\prime}_{m}\}$ as the jump and 
$\hat{H}^{\prime}_{\rm eff} =  \hat{H}^{\prime} - i \sum_{m=1}^M \hat{c}^{\prime}_m{} \dg \hat{c}^{\prime}_m /2$ as the no-jump operators, different $\vec{\beta}$ and ${\bm S}$ thus correspond to different measurement schemes~\cite{CarQTraj,WisMil10}.  See endnote~\cite{schemeComment} for more details.

Monitoring schemes can be divided into those that lead to jumps in the quantum state \cite{PhysRevLett.68.580,PhysRevA.46.4363,barchielli1993stochastic} and those that lead to quantum diffusion~\cite{CarQTraj,0305-4470-25-21-023,PhysRevA.47.1652}.  The distinction is that, in an infinitesimal interval of duration time $dt$, the former involves detector `clicks' that occur with a probability proportional to $dt$ and deliver a finite amount of new system information, whereas the latter always deliver an infinitesimal amount of information in this infinitesimal interval.  Here, we are concerned with quantum jumps, rather than diffusion, as this allows transitions analogous to those contained in the Einstein model.  However, there is typically still a large distinction between classical and quantum jump trajectories: in between the quantum jumps the experimentalist is continuously updating the system state in a non-trivial way, as the `no-click' results also carry information, albeit an amount that scales with $dt$.  This information affects the state even when it is a state of maximal information (i.e.~pure), unlike the classical case, leading to smooth but non-unitary evolution between jumps. Thus, it is clear that the system generically explores a continuum of states in Hilbert space.  

Whilst a non-zero dimensional manifold of states is therefore typically associated with continuous measurement, the \sch-Hughston-Jozsa-Wootters (S-HJW) theorem~\cite{SchPCP36,hughston1993complete,KirFPL06}, by contrast, gives physical meaning to any pure state ensemble representing 
a mixed state matrix (also called a density matrix) $\rho$. In particular, for finite $D={\rm rank}(\rho)$, one may consider ensembles,
\beq
\rho=\sum^{K}_{k=1}\wp_k\ket{\phi_k}\bra{\phi_k},
\label{rhoEnsemble}
\eeq 
for any choice of $K$, provided that $D\leq K<\infty$. 
The S-HJW theorem states that if there exists physically a purification, in a higher dimensional Hilbert space, of a system in a mixed state $\rho$, then for any ensemble that represents $\rho$, there is a way to measure the environment(s) --- that is, make measurements in the larger Hilbert space that act as the identity on the system Hilbert space --- such as to collapse the system  into one of the pure states $\ket{\phi_k}$ with the appropriate probability $\wp_k$.  Note that in general these states are not mutually orthogonal, even for $K=D$, and this must be so for $K>D$. 

The S-HJW theorem applies to a measurement on the environment at a particular time.  If this is a time remote from the initial conditions, and the system 
obeys \erf{me1} with a unique  stationary solution $\rho_{\rm ss}$ of rank $D$~\cite{rankMEcomment},  then in \erf{rhoEnsemble}, 
$\rho = \rho_{\rm ss}$. An obvious question is: can the finite ensembles representing $\rho_{\rm ss}$ allowed by the S-HJW theorem also pertain, at remote times, to continuous monitoring? To address this we make the additional assumption, mentioned above, that the ME has been derived from a QWN coupling.  Then we can ask whether a given pure state ensemble can be realised continuously by the experimentalist via a carefully chosen measurement scheme.  That is, is it possible, merely by obtaining information from the bath in the right way, 
to force a quantum system,  obeying a given ME, to behave like a discrete classical system, in the sense of jumping between a given finite set of pure states?  It was shown in Ref.~\cite{wiseman2001inequivalence} that this question is equivalent to asking whether the following
finite set of algebraic constraints can be satisfied 
\beq
\label{jumpCond} 
  \forall k, \ {\cal L}\ket{\phi_k}\bra{\phi_k} = \sum_{j=1}^K \kappa_{jk} 
 \left(\ket{\phi_j} \bra{\phi_j}-\ket{\phi_k} \bra{\phi_k} \right)
\eeq
for some ensemble $\{ \wp_k,\, \ket{\phi_k} \}$ of size $K$.  The real-valued transition rates, $\kappa_{jk} \geq  0$, naturally determine the occupation probabilities $\wp_k$.  A valid solution is known as a {\it physically realisable ensemble} (PRE)~\cite{wiseman2001inequivalence}, because there exists some measurement procedure that will realize the ensemble in the sense described above, even if that procedure may by difficult to implement in practice.  In particular, it is known that the measurement scheme required to achieve a PRE is generally adaptive in nature~\cite{karasik2011many}. In general, a ME will allow multiple solutions to \erf{jumpCond} via the experimental freedom described by \erfs{jumpOps}{noJumpOps}.  Most of the difficulty of our research program arises due to the system of non-linear constraints defined by \erf{jumpCond} being difficult to solve, even numerically, when $D>2$.

It was also shown in Ref.~\cite{wiseman2001inequivalence} that there are 
ensembles that represent $\rho_{\rm ss}$ but that are not PREs (this was referred to as the ``preferred ensemble fact'').  A fundamental question for open quantum systems is whether, for a given master equation, there exist {\em any} finite PREs. It was found in \crf{karasik2011many} that for $D=2$ it always possible to find at least one $K=2$ PRE. For $D>2$, a heuristic argument, 
 using free parameter and constraint counting, was made in~\crf{karasik2011many} predicting that one can expect a PRE to exist 
 if $K\geq (D-1)^2+1$. This separation from the classical case (where $K=D$ is necessary and sufficient) for $D>2$ would indicate a profound difference between quantum and classical open systems.  However, the heuristic argument of Ref.~\cite{karasik2011many} was not tested against numerical evidence, and both the quantum--classical gap, and the very existence of finite quantum ensembles in general, remained conjectural.  The question of whether ME symmetries can alter our expectations regarding the minimal size of PREs was treated in \cite{warwis2018a}.  There it was found that a commonly employed
% \pw{Why did you want to add this?  It seemed to me that the form of the invariant subspace symmetry that we used was actually less commonly used than the reduced Hilbert space dimension invariant subspace symmetry}
invariant subspace symmetry can reduce the heuristic ensemble size to $K\geq \frac{1}{2}\left(D^{2}-D+2\right)$, which is still larger than $D$ for $D>2$.

In this paper we address the three most important open questions raised by Ref.~\cite{karasik2011many}. We answer the first two definitively, and provide strong numerical evidence to support our conjectures regarding the third.  
The first question (Q1) is: {\it are there MEs for which the minimally sized PRE is larger than $D$?}  
We answer this in the positive by exhibiting a ME for which there are provably no PREs of size $D$. In this sense we can state (consistent with our title) that open quantum systems 
can be harder to track than open classical systems.  
The second question (Q2) is as follows: {\it is an ensemble size of $K=(D-1)^2+1$ always sufficient for a PRE to be found?}  
This time we will answer in the negative by exhibiting a ME for which there are provably no PREs of size $(D-1)^2+1$ or smaller. This proof requires the refining of the counting arguments from Ref.~\cite{karasik2011many}, which also leads us to a modified heuristic for the minimum sufficient $K$, containing a dependence upon the number of decoherence channels modeled by the ME. 
The third question (Q3) is: {\it does this refined form of the argument in Ref.~\cite{karasik2011many} reliably predict whether PREs are feasible for a ME of a given form}. 

For clarity, we partition Q3 into its natural sub-questions (Q3a, Q3b, Q3c), arising from the heuristic's comparison of the number of free parameters and constraints that constitute the algebraic formulation of PREs.  Q3a asks whether the heuristic's prediction of ruling out PREs for ensembles smaller than the determined threshold is accurate, while Q3b assesses its utility when the number of parameters and constraints are equal and, finally, Q3c concerns scenarios where there are more parameters than constraints. The heuristic can make predictions for MEs and PREs in particular classes, so we take the prediction of the heuristic to be as follows: 
only in a measure-zero set of MEs will PREs exist contrary to expectation, whilst when PREs {\it are} feasible according to the heuristic, they will exist for a finite fraction of randomly drawn MEs.  The numeric results that we obtain strongly support the conjecture that the counting heuristic makes accurate predictions for Q3a and Q3b, in the sense described. 
At the same time, the numerics undermine the unreasonably strong  hypothesis 
that the heuristic is a necessary and sufficient condition for the existence of PREs.  Some discussion of Q3c is provided in our concluding remarks, but a systemic investigation is left for future work due to its somewhat divergent focus from Q1-2 and, also, its computational difficulty.

It is worth highlighting the reasons why Q1-Q3 have not been previously answered and, in turn, why we are now in a position to answer them. The most relevant point is that PREs, in $D>2$, are {\it hard} to investigate.  There are no known analytic expressions for their construction and, more importantly, even numerically their discovery is extremely difficult.  To find a PRE, the set of nonlinear polynomial constraints given by \erf{jumpCond} must be solved.  The difficulty of this task becomes exponentially more difficult as the number of equations and variables increases.  In fact, the problem is known to fall into the NP-complete complexity class~\cite{courtois2000efficient}.  This alone does not prohibit the constraints' solution; it just places low practical bounds on the system size that can be solved.  Thus, with regards to Q1-Q3, it becomes an issue of how large a system can be numerically processed in comparison to how small a system is minimally sufficient for investigation.  In this paper we utilise the recent work regarding symmetries for PREs that was carried out in \cite{warwis2018a}.  This allows us to search, with meaningful expectation, for PREs possessing symmetry that are of a smaller size than that expected based on the heuristic of \cite{karasik2011many}.  Computationally, the effect of this is to reduce the size of the relevant polynomial system to only 24 constraint equations, as compared to the 45 constraint equations that are relevant when simplifying symmetries are not considered.  In terms of enlarging the size of the polynomial systems that we can address, we newly apply two powerful software packages (MAGMA~\cite{bosma1997magma} and PHCpack~\cite{verschelde1999algorithm}) to PREs that respectively take advantage of Gr{\"o}bner basis~\cite{cox2006using} and polynomial homotopy continuation~\cite{li1997numerical} techniques.  In this way, by applying symmetry and more powerful numerics, we are able to cross into the regime where Q1-Q3 can be answered.

The paper is organised as follows.  Firstly, in \srf{symmPREs}, a brief description of the symmetries introduced in \cite{warwis2018a} is provided.  In \srf{countArguments} we then develop a more sophisticated heuristic than that found in \crf{karasik2011many} to guide when PREs are expected to exist and when not to exist.  This provides a pathway to follow in order to answer the research questions Q1--Q3 introduced above.  In \srf{computationalDifficulty}, the mathematical task of solving \erf{jumpCond} is discussed.  In \srf{results}, we address Q1--Q3 using state-of-the-art numerical, and computational algebra, solvers for various classes of MEs and PREs. Finally, we conclude with a summary and a discussion of future research directions in \srf{conc}.

\section{Symmetry and PREs}
\label{symmPREs}
In prior work, the role of various types of symmetry in the structure of PREs was explored~\cite{warwis2018a}.  
It was found that symmetries of the ME can be used to simplify the task of finding PREs and, also, to relate discovered PREs.  That is, PREs can inherit ME symmetries in a well defined way.  In this current paper, we will utilise symmetry for systems of dimension $D=3$, in order to facilitate obtaining answers to the research questions posed in the introduction.  For completeness, we now provide a brief simplified description of the symmetry tools that were developed in \cite{warwis2018a}.

\subsection{Invariant subspaces of ${\cal L}$}
\label{invSubSec}
We label the space of density matrices for a $D$-dimensional complex Hilbert space $\mathbb{H}$
as $\mathfrak{D}\left({ \mathbb{H}}\right)$.  We say that a ME, $\dot{\rho}={\cal L}\rho$, has an invariant subspace symmetry iff there exists some non-trivial subregion, $\mathfrak{D}_\mathfrak{I}$, of $\mathfrak{D}\left({ \mathbb{H}}\right)$, to which dynamics 
%(described by evolution operator $e^{{\cal L}t}$) 
is confined, given an initialisation within that subregion.  Additionally, we require that the subregion be an interesting one, in the sense that it has the potential to support PREs for the specified ME.  This latter point requires that it contain at least $D$ pure states, in order to form pure-state ensembles of rank $D$ --- we have stipulated that $\rho_{\rm ss}$ be of rank $D$. An example of such an invariant subspace, provided in \cite{warwis2018a} and of relevance here, is that of real-valued density matrices --- otherwise known as `redit' states.

The reason for considering such invariant subspaces is that one can then look for solutions to \erf{jumpCond} that lie entirely in this subspace.  That is, 
\beq
  \forall k,\quad \ket{\phi_k}\bra{\phi_k} \in\mathfrak{D}_\mathfrak{I}.
\label{invSubspace}
\eeq
One can consider \erf{jumpCond} as effectively enforcing constraints relating to the different regions of $\mathfrak{D}\left({ \mathbb{H}}\right)$: given that $\ket{\phi_k}\bra{\phi_k}$ does not have support on all such regions, some portion of the constraints are automatically satisfied.  The reduction in the number of non-trivial constraints upon $\ket{\phi_k}\bra{\phi_k}$ means that the polynomial system that must be solved in order to find PREs (of the form \erf{invSubspace}) is smaller than if an invariant subspace was not considered.  Additionally, there is a potential reduction in the expected minimum PRE size.  This will be discussed further from a theoretical perspective in \srf{countArguments} and be utilised in \srf{results}, where we find new PREs.  Note that in performing a limited search for PREs, one is excluding the possibility of finding the fraction (which may be zero, one or in between) of PREs that have support outside $\mathfrak{D}_\mathfrak{I}$.

\subsection{Wigner symmetries}
\label{wigSymmSec}
The second ME symmetry discussed in \cite{warwis2018a} that is of direct relevance, is that of Wigner symmetry.  Wigner transformations act  on Hilbert space rays in a way that preserves the Hilbert space inner product, $|\braket{\psi_{1}}{\psi_{2}}|$. Their action is consequently well defined upon pure state projectors, and we denote this as ${\cal T}\ketbra{\psi}{\psi}$. Wigner showed that such transformations are either unitary (and so linear in their action on Hilbert space) or antiunitary (and so antilinear)~\cite{wigner1,wigner2}.  In this subsection, we are concerned with those Wigner transformations that leave the Lindbladian, ${\cal L}$, invariant:
\beq
{\cal T}^{-1}{\cal L}\ {\cal T}={\cal L},
\label{Uinv}
\eeq
and term these `Wigner symmetries'. 

Given a ME whose Lindbladian satisfies \erf{Uinv} for some ${\cal T}$, we can then posit the existence of PREs that possess the Wigner symmetry, which we define in the following way.
Let $P$ be a permutation of $\{1...K\}$ with $k^{\prime}=P(k)$.  Then we define a PRE as having the Wigner symmetry ${\cal T}$ iff $\exists$ permutation $P$ such that
\bqa
\forall k, \ {\cal T}\ket{\phi_k}\bra{\phi_k}& =& \ket{\phi_{k^{\prime}}}\bra{\phi_{k^{\prime}}}\quad\quad{\rm and}
\label{induced}\\
 \kappa_{j^{\prime}k^{\prime}} &=&\kappa _{jk}.
\label{inducedMap}
\eqa
The consequence of both the ME (through ${\cal L}$) and the PRE possessing the Wigner symmetry [\erf{Uinv} and \erfs{induced}{inducedMap}, respectively] is that some portion of the constraints of \erf{jumpCond} are redundant~\cite{warwis2018a}.  Specifically, given an equivalence relation, $\sim$, amongst ensemble members in the presence of the symmetry ${\cal T}$ as 
\beq
\ket{\phi_k}\sim\ket{\phi_{k^{\prime}}} \quad {\rm iff} 
\quad 
\exists \ \  {\cal T}: \ %\quad
\ket{\phi_{k^{\prime}}}\bra{\phi_{k^{\prime}}}=
{\cal T}\ket{\phi_{k}}\bra{\phi_{k}},
\label{equivClass}
\eeq
then the constraints on only one element of each equivalence class, $[\ket{\phi_k}]_{\sim}$, 
need to be tested as the remainder are implied.  Once again, the benefit is a smaller polynomial system that must be solved in order to find PREs.  Similarly to the invariant subspace symmetry, Wigner symmetry will be used in \srf{results} to find new PREs.

\section{Heuristics for the existence of PREs}
\label{countArguments}
Previous work~\cite{karasik2011many} argued heuristically, via the counting of free parameters and constraints of \erf{jumpCond}, that typically $K\geq(D-1)^2+1$ ensemble states are required for a PRE to be possible.  More recently~\cite{warwis2018a}, consideration was given to the minimum PRE size in the presence of the invariant subspace symmetry $\mathfrak{D}_\mathfrak{I}=\R^{D\times D}\cap\mathfrak{D}\left({ \mathbb{H}}\right)$ (real-valued density matrices), and an analogous heuristic, $K\geq \frac{1}{2}\left(D^{2}-D+2\right)$, was derived.
Such a $\mathfrak{D}_\mathfrak{I}$, containing redit states, provides an example of a subregion that has the minimal dimensionality (when considered as a convex linear space) able to support $\rho_{{\rm ss}}$ having ${\rm Rank}(\rho_{{\rm ss}})=D$, as is appropriate for the minimisation of $K$.

%More recently~\cite{warwis2018a}, consideration was given to the minimum PRE size in the presence of an invariant subspace symmetry, and an analogous heuristic derived ($K\geq \frac{1}{2}\left(D^{2}-D+2\right)$ in the case that the subspace in minimally sized, as is appropriate for the minimisation of $K$).

The existence of free parameters in \erf{jumpCond} is due to there being no preference (at least none relevant to our current discussion) as to the nature of the states, $\ket{\phi_k}$, comprise the PRE, nor of the transition rates, $\kappa_{jk}$, between them.  The \blk constraints obviously enforce the properties intrinsic to the PRE.  That larger $K$ makes it more likely for a PRE to exist, all other things being equal, arises due to the number of free parameters (we will often omit the modifier `free' going forward) depending quadratically upon $K$ while the number of constraints only depends on it linearly. (The origin of these dependences will be discussed in detail later.)  

An important piece of evidence supporting the argument in \cite{karasik2011many} for $K\geq(D-1)^2+1$ is the fact (proven in \cite{karasik2011many}) that every system obeying a Lindblad-form master equation has at least one PRE with $K=2$ for $D=2$.  For $D,K=2$ it is indeed the case that the number of parameters and constraints involved in \erf{jumpCond} are equal (we term this a `square' system).  Moreover, the solution set in that case typically \blk consists of isolated solutions (termed a `zero-dimensional solution set'), consistent with the equality portion of $K\geq(D-1)^2+1$ holding.  Similar remarks apply in the case of a minimally dimensioned \blk invariant subspace, such as that of rebit states \blk [$(D-1)^2+1=\frac{1}{2}\left(D^{2}-D+2\right)$ for $D=2$].

In this section a more sophisticated heuristic for the required ensemble size $K$ is developed that has dependence not only upon $D$ but also 
upon the number of decoherence channels, $L$ in \erf{me1}. \blk There is, of course, no guarantee that a physical solution (representing a PRE) can be found when the number of constraints is less than or equal to the number of parameters but it serves as a heuristic for when one is likely to find PREs. This heuristic is important even for a computational approach to the existence problem, given the computational complexity of finding PREs (discussed in detail in \srf{computationalDifficulty}).

The description of a PRE consists of a set of states and their occupation probabilities, but
when investigating their existence it is more more beneficial to think about the transition rates, $\kappa_{jk}$, rather than $\wp_k$.  This is because the structure of the ME can (in a way that is defined in \srf{rateCounting}) place constraints on the allowed $\kappa_{jk}$, which can simplify \erf{jumpCond}.  To aid these arguments it is useful to form a graph representation of the ensemble in which each member (state) is a node and allowed transitions are illustrated by directed edges.  An example is shown in \frf{graphs}(a) for 4 ensemble members with all transitions allowed (termed `fully connected').  It is possible that some of the allowed transitions are, in fact, not utilized ($\kappa_{jk}=0$ for some $j,k$) by a particular ensemble.

 \begin{figure} [t]
\centering
\includegraphics[scale=0.5]{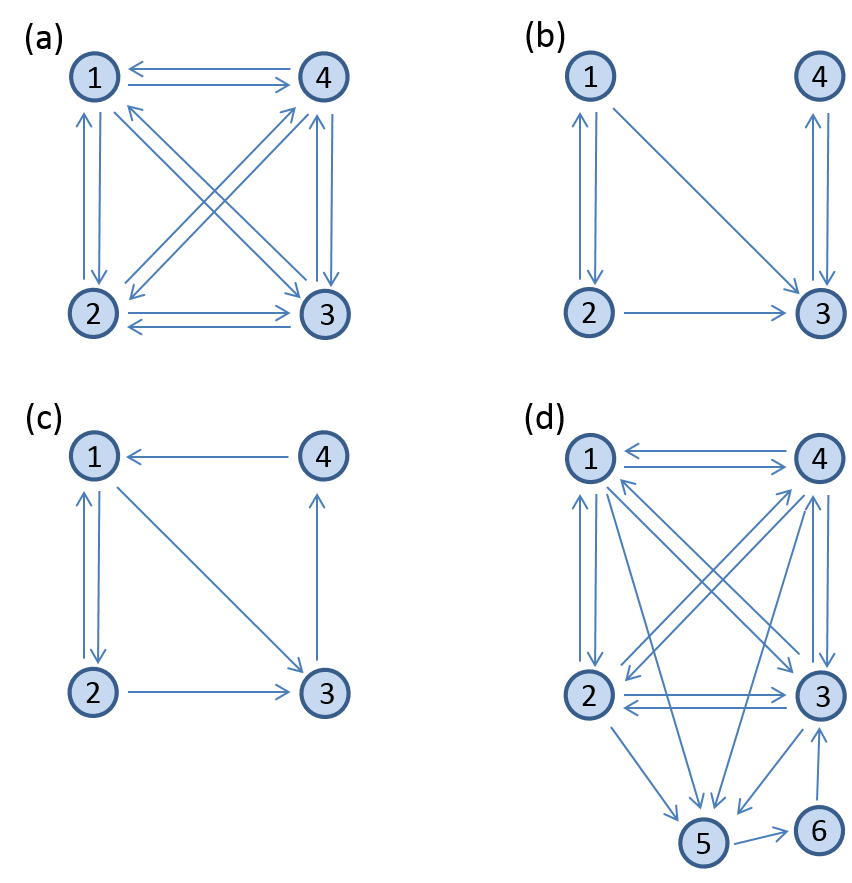} 
\vspace{-2mm}
\caption{Each node represents a member state of a PRE and directed edges are transitions in the direction of the arrow. (a) A 4 member fully connected graph.  (b) A graph that traps the state in nodes 3 and 4 so is not a feasible PRE.  (c) Maximum number of transitions for a $D=3,L=1$ ME for a $K=4$ PRE.  (d)  Maximum number of transitions for a $D=3,L=1$ ME for a $K=6$ PRE. 
}
\label{graphs}
\end{figure}

\subsection{Parameter and constraint counting}
\label{rateCounting}
In this subsection we first form counting arguments for the arrangement and number of transitions that can exist, given the ME defining the system dynamics.  At the end of the subsection the more straightforward task of describing the number of state vector parameters, and the number of constraints upon them, is performed.  

It is of interest to maximise the number of graph transitions as this will give the greatest chance of PRE existence, at least according to our heuristic criterion for which the transitions exist as free parameters. 
The graph that maximises the number of transitions is obviously the fully connected one but, as will be shown, this will not always be consistent with the ME in question.  Perhaps the most obvious requirement of a graph --- and one that the fully connected graph satisfies --- is that all the nodes must be repeatedly explored in the long time limit.  That is, the state cannot become trapped in a sub-graph.  An example of a graph that is ruled out by this consideration is \frf{graphs}(b).  We note that freedom in choosing the measurement scheme is assumed; in particular a sufficient number of detectors, $M$, is provided to avoid unnecessarily limiting the number of possible different post-jump states.   

Let us now discuss the circumstances under which we can rule out the fully connected graph.  To do so in a simple manner, we will consider Hilbert subspaces, $\tilde{\mathbb{H}}$, as opposed to subregions of $\mathfrak{D}\left({ \mathbb{H}}\right)$.  
That is $\mathbb {H}= \tilde{\mathbb{H}}\oplus{\mathbb{H}_{R}}$ (with, $R$ being the remainder space).  In the case that $\mathbb{H}_{R}$ is non-trivial, then dim$(\tilde{\mathbb{H}})<D$. In this subsection, we will often refer to the dimension of an ensemble of pure states, by which we mean the dimension of the Hilbert subspace, $\tilde{\mathbb{H}}$ that contains the entire ensemble. 
%It is important to realize that our notion of invariance is different from that of Ref.~\cite{4639467}.  We consider invariant subspaces of $\mathfrak{D}\left({ \mathbb{H}}\right)$, 
%with dim$(\mathbb{H})=D$.
%We require that the invariant space contains rank $D$ density matrices (meaning that there will exist density matrices that are formed by an ensemble of $D$ linearly independent pure state projectors).
%This is in contrast to Ref.~\cite{4639467}, where the invariant subspace, of Hilbert space dimension $\tilde{D}$, is $\mathfrak{D}( \tilde{\mathbb{H}})$  for $\mathbb{H}= \tilde{\mathbb{H}}\oplus
%{\mathbb{H}_{R}}
%$ (with, here, $R$ being the remainder space).  In other words, their invariant space consists of all possible density matrices with support solely in some Hilbert subsystem, $\tilde{\mathbb{H}}$, whereas that considered in this paper is a compact subregion $\mathfrak{D}_{\mathfrak{I}}\subset\mathfrak{D}\left( \mathbb{H}\right)$.
%Note that in the long time limit, under ME dynamics, the state necessarily relaxes to the unique $\rho_{{\rm ss}}$, a single point within the invariant subspace.
Each directed connection (with rate $\kappa_{jk}$) references the transition from the $k$th to the $j$th state, which will occur when a detection event is registered at a, possibly non-unique, detector, $m$ such that $f(m,k)=j$. 
The function $f$ takes as inputs the clicking detector, $m$, and the pre-click system state $k$, to give a post-click state $j$ (see \arf{finMeasScheme} for more details concerning PRE measurement schemes).  We are interested in the case $j\neq k$, so that the detection causes 
a finite change in the state, from $\ket{\phi_k}$ to $\hat{c}_{m}^{k}\ket{\phi_k}\propto\ket{\phi_{f(m,k)}}$, where $\hat{c}_{m}^{k}$ is $m$th jump operator when the system is known (or, rather, believed) to be in the $k$th state.  Although there can be many different possible post-jump states $j$ from a particular state $k$, the dimension of the post-jump Hilbert subspace (which plays the role of the previously defined $\tilde{\mathbb{H}}$) is restricted in size to be $\leq\min \{L+1,D\}$.  This is because the $\hat{c}_{m}^{k}$ are formed from linear combinations of $\hat{c}_{l}$ (of which there are $L$) together with the identity (see \erf{noJumpOps} and \arf{finMeasScheme}), 
while $D$ is the dimensionality of the system. In the case where this restriction is saturated, the pre-jump state, $\ket{{\phi_k}}$, also belongs to $\tilde{\mathbb{H}}$.

The importance of this restriction can be seen by examining a $K=4$ ensemble.  First we assume that the rank of $\rho_{\rm ss}$ is $D=3$ and that the number of Linblad operators is $L=1$.  
For any state in the ensemble, the Hilbert subspace comprising this state and the post-jump state(s), say $\tilde{\mathbb{H}}^{\prime}$, must be of dimension 2. If all the transition rates were non-zero then 
$\tilde{\mathbb{H}}^{\prime}$ would contain the entire ensemble, which is a contradiction as 
dim($\tilde{\mathbb{H}}^{\prime})<D$.  Thus, the fully connected graph of \frf{graphs}(a) is ruled out as a possibility for $L=1$ and the number of free parameters corresponding to transition rates is less than the expected 
$K(K-1) = 12$.  By contrast, if the rank of $\rho_{\rm ss}$ is kept at 3 but $L=2$ then 
dim$(\tilde{\mathbb{H}}^{\prime})$ can be 3 and the fully connected graph is not ruled out on these grounds.  

Coming back to ${\rm Rank}(\rho_{\rm ss})=3$ and $L=1$, the obvious question is: what is the largest number of transitions allowed?  An example optimal configuration is given in \frf{graphs}(c), where 6 rates are possible.  Nodes $1,2,3$ do span a 2-dimensional Hilbert subspace but there is asymmetry as node $4$ is only accessed via node 3's single outward connection.  This conspires to allow node $4$ to be outside the node $1,2,3$ subspace and increase the Hilbert space \blk dimension of the ensemble to 3 as required.  That only 6 rates are possible represents a large reduction from the fully connected graph.  
It will be shown that $K=4$ PREs are not generically allowed for $D=3$ for any $L$.  However, MEs possessing certain types of symmetry can render some of the constraints of \erf{jumpCond} redundant. Together with introducing such symmetries, if we take $L\geq2$ (so that fully connected graphs are not ruled out by the considerations of dim$(\tilde{\mathbb{H}}^{\prime})$ that are discussed above), then the resultant system of polynomial constraints becomes `square' and thus PREs may be searched for with significant expectation of their existence. \blk
%{This will allow the creation of a square system of equations for $L\geq 2$, where the above mentioned constraints on the number of rates for $L=1$ are irrelevant. }

Another example ($K=6, L=1,D=3$) that will prove relevant and that also serves to illustrate the optimal technique for `rate packing' is shown in \frf{graphs}(d).  The idea is that there is as large a group of highly connected nodes as possible (nodes $1$--$4$ in \frf{graphs}(d)), because this allows the number of connections to grow as the square of the group size.  As the dimension of the ensemble is required to reach $D$, there is an outward connector node that breaks the symmetry, here node  5.  Node 5 cannot be connected to anything in addition to 6 as, if it were, this would constrain node 6 to be in the same subspace as nodes 1--4.  Node 6 further breaks the symmetry by connecting to one (and only one) of the nodes 1--4.  A single connection from 6 to 5 is not allowed as this would trap the system in nodes 5,6.  If, instead of $K=6$, the ensemble size was $K=7$, the extra node would be optimally added to the highly connected group, with then $9$ more transitions possible.  If it was added anywhere else fewer connections would be possible.  As a different extension, if $D=4$ then one of the highly connected nodes would have to be moved to the lowly connected chain that serves to increase the Hilbert space dimension of the ensemble; for example, in between nodes 5 and 6. 

We now have enough intuition to give the general case, describing how the maximal number of transition rates can be packed for arbitrary $K,D,L$.  The schematic for the configuration is given in \frf{ratePack} and is based on the principle of a highly connected group (of dimension $L+1$) that contains most of the transitions and a lowly connected group whose purpose is to fill out the ensemble dimension to match Rank($\rho_{\rm ss}$).  There are a few generalisations to the $L=1$ examples above.  Firstly, the node which connects the upper to lower groups is also connected internally within the upper group, but less so than the other upper members.  As long as it has no more than $L$ connections, there is no effect on the upper and lower dimensionalities.  For the case $L=1$ there is only one connection so this is consistent with the analysis above.  The second major difference is that the lower group is also internally connected up to a maximum of $L$ connections, by the same reasoning as just given.  The total number of transition rates for arbitrary $K,D,L$ can easily be counted from \frf{ratePack} to give $(K-D+L)^{2}+L(D-L)$.  This expression applies when $L<D-1$.  If $L\geq D-1$ there is no need for a lowly connected group of nodes and the graph can be fully connected, with $K(K-1)$ rates.  

Before comparing the {\it total} number of parameters and constraints, in \srf{re3it}, we first give the number of parameters in \erf{jumpCond} that arise from the specification of the $K$ ensemble members.  A generic $D$-dimensional pure state can be described by a state vector with $D$ complex numbers or $2D$ real numbers, but this is before normalisation and removing an overall phase.
Thus, $2D-2$ real parameters per state are sufficient, or $2K(D-1)$ in total.  Finally, to compare against the number of parameters, we need the number of constraint equations.  To find this, we note that both sides of \erf{jumpCond} are, by construction, Hermitian and traceless, which leads to $D^2-1$ constraints for each $k$.

% \subsubsection{Parameter and constraint counting for $\mathfrak{D}_\mathfrak{I}=\R^{D\times D}\cap\mathfrak{D}\left({ \mathbb{H}}\right)$}
Finally in this subsubsection, we comment on the special case of the invariant subspace symmetry $\mathfrak{D}_\mathfrak{I}=\R^{D\times D}\cap\mathfrak{D}\left({ \mathbb{H}}\right)$, which we use extensively in the latter parts of this paper. 
In this case, only $D-1$ parameters are required to describe each state vector.  Any smaller invariant subspace would not support rank$(\rho_{{\rm ss}})=D$.  The number of non-trivial constraint equations is also reduced in number, to $(D^2+D)/2$ per ensemble member, as any constraints relating to the imaginary components of the state vector are trivially satisfied.
%\pw{Don't think I agree that this should be moved to 3.2.1 or 3.3.  Section 3.2 requires knowledge of the number of parameters so this paragraph needs to be before (or in) that section. The current section is titled parameter and constraint counting, and that's what's being done, albeit for a special case.  If anything, I would say 3.1.1 (which is what I've tentatively done.}

\subsubsection{Ambiguity in the counting of state parameters}
\label{dimCounting}

As \erf{jumpCond} is a matrix equation it would be natural to consider forming the ensemble members as state {\it matrices} rather than state vectors.  If this approach is taken, then the counting is different as follows.  A $D$-dimensional Hermitian matrix with unit trace is given by $D^2-1$ real parameters.  To ensure that the state is pure and is positive, the additional (real) constraints $\Tr\left[\rho^2\right]=\Tr\left[\rho^3\right]=1$ need to be applied~\cite{schumacher2010quantum}.  On the face of it this removes 2 free parameters.  However, we quickly run into difficulties.  For example, if $D^2-3$ real parameters are used for a pure state matrix then this gives 6 free parameters for $D=3$ rather than 4 via a state vector approach.  Thus, the state matrix description is not minimalistic in the number of parameters and only becomes equivalent to the state vector description with the highly non-linear cubic constraints imposed by $\Tr\left[\rho^3\right]=1$.  We posit that the parameter and constraint counting {\it should} be conducted in a minimal way and consequently proceed in that direction.  Evidence for the correctness of this approach is subsequently found and described further in \srf{majorQ}, where we establish that no PREs are possible for specific MEs when the number of parameters is less than constraints as determined by the state vector approach but the number of parameters is larger than constraints with the matrix method.

 \begin{figure} [t]
\centering
\includegraphics[scale=0.5]{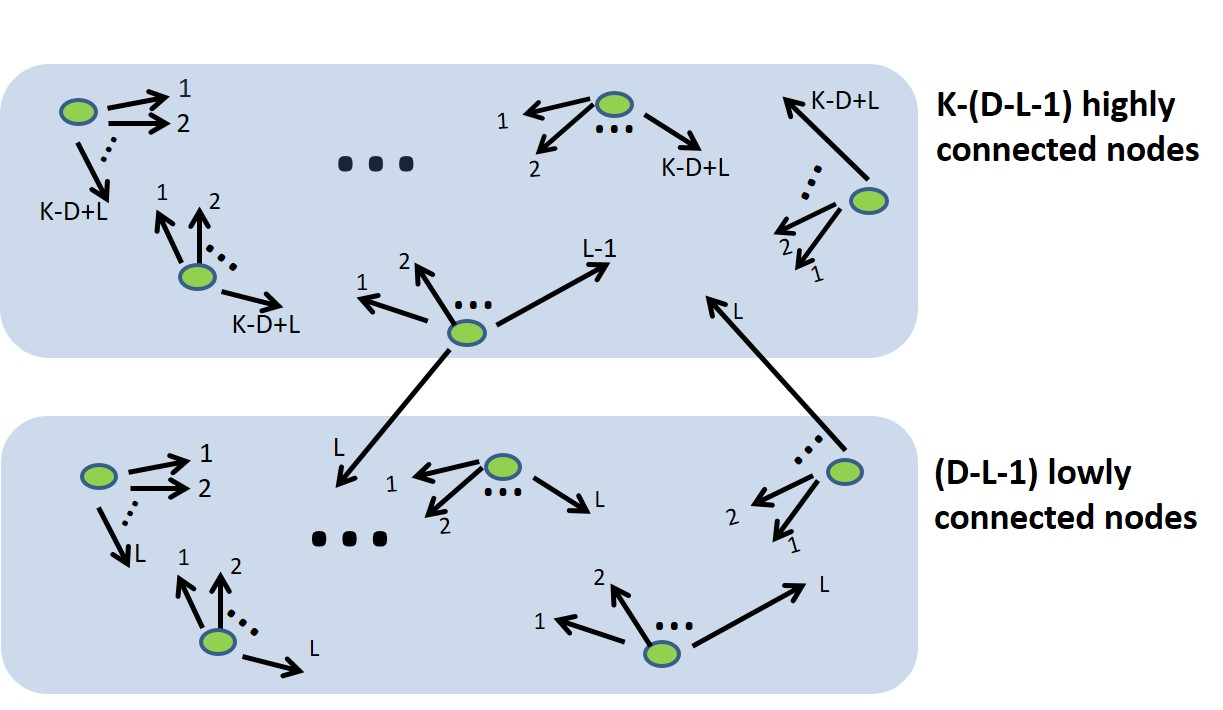} 
\vspace{0mm}
\caption{Schematically shown is the heuristic for optimal rate packing in a PRE.  The ensemble states are divided into two classes, the first of which (upper panel) is highly connected and the second (lower panel) is less connected and serves to increase the ensemble dimension.  The highly connected nodes are fully intra-connected with the exception of the symmetry breaking node that connects to the less highly connected nodes.  The dimension of the highly connected nodes is $L+1$.  The lowly connected nodes fill out the dimension of the ensemble to $D$, with each node increasing the dimension by one.  This is achieved by having no more than $L$ connections from any one node.  Finally, the system state must not be trapped in the lower sub-section, so there is at least one connection from lower to upper.
 }
\label{ratePack}
\vspace{-4mm}
\end{figure}

\subsection{Comparing the number of parameters and constraints}
\label{re3it}
As a reminder to the reader, we have adopted the following heuristic when searching for the PREs: the total number of free parameters should be greater than or equal to the number of constraints as described by \erf{jumpCond}.  Requiring this to hold for a $K$-sized ensemble in $D$ dimensions with $L<D-1$ Lindblad operators (in the absence of invariant subspace symmetry) thus gives 
\beq
(K-D+L)^{2}+L(D-L)+2K(D-1)\geq K(D^2-1).
\label{PREineq}
\eeq 
Here we have not simplified the inequality so that the number of transitions (terms one and two), the number of state parameters (term three) and the number of constraints (RHS) can be clearly identified.
If $L\geq D-1$ then the relevant inequality is that obtained in~\crf{karasik2011many}: $K\geq(D-1)^2+1$.  The minimum ensemble size, $K_{{\rm min}}$, can be found from \erf{PREineq} for fixed $D, L$ by taking the smallest integer value of $K$ that satisfies it. 
The integer rounding associated with $K_{{\rm min}}$ simplifies the quadratic solution in such a way that we can summarise our knowledge of the minimum expected PRE size, for all $D,L$, as:
\beq
K_{{\rm min}}= (D-1)^{2}+1+{\bf 1}_{\{L< D-1\}}\times \left(2D-2L-1 \right),
\label{simpleKMin}
\eeq 
where ${\bf 1}_{\{A\}}$ is the indicator function, which is 1 if $A$ is true and 0 otherwise. 
That is, if $L \geq D-1$ then we reproduce the minimum ensemble size suggested in \crf{karasik2011many}. But if $L<D-1$, the minimum ensemble size is larger by 
$2D-2L-1$, making it equal to $K_{{\rm min}}=D^2-2L+1$.  For all values of $L$, $K_{{\rm min}} \sim D^2$ and in the `worst' case, of $L=1$, we find $K_{\rm min} =  D^2-1$ (for $D>2$). 
The values of $K_{{\rm min}}$ for small values of $D,L$ are summarized in \trf{table1}.

Equipped with the expected ensemble size, $K_{{\rm min}}$, required for a PRE to exist, we can  assess which classes of MEs are appropriate in order to address the targetted research questions (Q1-3) that were raised in \srf{intro}.  Firstly, we consider when an ensemble of size $K_{{\rm min}}>D$ is required, which relates to Q1.  From \trf{table1}, we expect that this will be when $D\geq 3$ and will be minimally satisfied with $K_{{\rm min}}=5$ for $L\geq 2$ and $D=3$.  Secondly, to answer Q2 we consider whether $K=(D-1)^2+1$ is always sufficient to find a PRE.  We have shown that, at least in terms of parameter and constraint counting, it is not (for $D>2$), as 
MEs with $L<D-1$ are expected to require an additional $2D-2L-1$ ensemble members.  
For example, a $D=3$ ME with a single decoherence channel has $K_{{\rm min}}=8$.
Whether we can actually consistently find PREs of size $K_{{\rm min}}$ (which relates to Q3b) and rule them out for $K<K_{{\rm min}}$ (Q3a) will provide evidence as to the quality of the heuristic and thus constitute our response to Q3.

\subsubsection{$K_{\rm min}$ for a minimally sized invariant subspace}
The analogous expressions to \erfs{PREineq}{simpleKMin} can be found when the invariant subspace symmetry is considered, specifically the space of real-valued density matrices.  That is, we find the minimum PRE size, given that it is restricted to the subspace and that the ME maps real-valued density matrices to real-valued density matrices.  Comparing the sum of the transition parameters and state parameters with the number of constraints gives, when $L<D-1$,
\beq
(K-D+L)^{2}+L(D-L)+K(D-1)\geq \frac{1}{2}K(D^2+D-2),
\label{PREineqREAL}
\eeq
while for $L\geq D-1$ it is $K(K-1)+K(D-1)\geq \frac{1}{2}K(D^2+D+2)$.  Solving this latter expression, for smallest integer $K$, one obtains $K_{{\rm min}}= \frac{1}{2}\left(D^{2}-D+2\right)$.  
The minimum PRE size for arbitrary $D,L$ can thus be written as
\begin{equation}
 K_{{\rm min}}=\frac{1}{2}\left(D^{2}-D+2\right)+{\bf 1}_{\{L< D-1\}}\times (2D-2L-1)-{\bf 1}_{\{g(D,L)\}}, 
\label{simpleKMinReal}
\end{equation}
with 
$g(D,L)$ being a Boolean function, whose dependence upon $D$ and $L$ 
is not particularly important as it only influences $K_{{\rm min}}$ by one unit. 
More interesting is the fact that, to within one unit, the addition to the ensemble size necessitated by small $L$ is precisely the same as in the generic (non-invariant subspace) case of \erf{simpleKMin}: $2D-2L-1$. Moreover, the threshold for when this addition applies is also the same: $L<D-1$. Example values for $K_{{\rm min}}$, for small $D,L$, are given in \trf{table2}. 

From \trf{table2}, we expect that the minimal ensemble size, $K_{{\rm min}}$, will be larger than $D$ for $D\geq 3$.  The lowest dimensioned such example is $D=3$ which has $K_{{\rm min}}=4$ when $L\geq 2$.  With regards to whether $K=(D-1)^2+1$ is always sufficient to find a PRE (research Q2), we see that even when an invariant subspace symmetry is applicable it is not expected to be sufficient for small $L$ (for example, $K_{{\rm min}}=6$ for $D=3$ and $L=1$).  In \srf{results} we will confirm the practical relevance of \trf{table2}.

In this subsection we have considered how $K_{{\rm min}}$ is modified for small $L$ when the invariant subspace symmetry is taken into account.  In \arf{wignerGraphsApp}, we provide some brief comments as to how another symmetry --- Wigner symmetry --- can impact on viable PRE graphs.  This will, in turn, effect the minimum size of PREs that possess the Wigner symmetry.

\begin{table}
 \centering

\begin{minipage}[t]{.45\textwidth}
%\begin{subtable}{.95\textwidth}
\centering

%\begin{tabular}{cccccc}
%\begin{tabularx}{0.95\textwidth}{c *{6}{Y}}
\begin{tabularx}{0.95\textwidth}{YY@{}Y@{}Y@{}Y@{}Y}

%\begin{tabular}{|*{6}{>{\centering\arraybackslash}p{.1\linewidth}|}}
%\hline
&\multicolumn{5}{c}{Lindblads} \\
\cline{2-6}
Dim.    & 1 & 2&3&4&5 \\
\hline
%1&1 &1&1&1&1 \\
2&2 & 2&2&2&2 \\
3&\cellcolor{gray!25}8  & 5&5&5&5 \\
4&\cellcolor{gray!25}15  & \cellcolor{gray!25}13 &10&10&10 \\
5&\cellcolor{gray!25}24  &\cellcolor{gray!25} 22 &\cellcolor{gray!25}20 &17&17 \\
\hline
\end{tabularx}

\caption{The minimum number of PRE members, $K_{{\rm min}}$, required for the number of parameters to equal or exceed the number of constraints is provided for a variety of $D,L$.  Our heuristic suggest that a soln of \erf{jumpCond} may be possible for $K=K_{{\rm min}}$.  Generic MEs are being considered.  The shaded cells highlight values of $K_{{\rm min}}$ that are made larger than $(D-1)^2+1$ (the ensemble size suggested in Ref.~\cite{karasik2011many}) due to the restrictions on allowed graphs described in \srf{rateCounting} (which apply when $L< D-1$).
}
\label{table1}
%\end{subtable}%
\end{minipage}\quad%
%
%\caption{blag}
%\label{table1}
%\end{table}
\begin{minipage}[t]{.45\textwidth}
%\begin{subtable}{.95\textwidth}
\centering
%\begin{tabular}{cccccc}
%\begin{tabularx}{0.95\textwidth}{c *{6}{Y}}
\begin{tabularx}{0.95\textwidth}{YY@{}Y@{}Y@{}Y@{}Y}

%\begin{tabular}{|*{6}{>{\centering\arraybackslash}p{.1\linewidth}|}}
%\hline
&\multicolumn{5}{c}{Lindblads} \\
\cline{2-6}
Dim.    & 1 & 2&3&4&5 \\
\hline
%1&1 &1&1&1&1 \\
2&2 & 2&2&2&2 \\
3&\cellcolor{gray!25}6  & 4&4&4&4 \\
4&\cellcolor{gray!25}11 &\cellcolor{gray!25} 10  &7&7&7 \\
5&\cellcolor{gray!25}17 &\cellcolor{gray!25} 15 &\cellcolor{gray!25}14 &11&11 \\
\hline
\end{tabularx}

\caption{The difference from \trf{table1} is that MEs with ${\cal L}$ having the redit as an invariant subspace are considered, as described in \srf{invSubSec}.  PRE states are restricted to being redits, which reduces the size of $K_{{\rm min}}$ necessary for the number of constraints imposed by \erf{jumpCond} to be less than, or equal to, the number of parameters.}
\label{table2}
%\end{subtable}
\end{minipage}
\end{table}

\section{Analysing polynomial constraints in order to find --- or rule out --- PREs}
\label{computationalDifficulty}
In our introductory comments, and also in previous work~\cite{warwis2018a}, we have commented on the difficulty of finding PREs.  In this section we provide further details, and also introduce the newly applied technique of polynomial homotopy continuation.  Additionally, the related task of definitively ruling out the existence of PREs of a particular size, for a specified ME, is discussed.

To progress beyond the heuristic arguments of \srf{countArguments}, example PREs need to be found for explicit MEs.  That is, the set of matrix equations specified by \erf{jumpCond} needs to be solved.  On the LHS of \erf{jumpCond}, there is a quadratic dependence upon the pure state vector parameterization (see~\srf{dimCounting}).  The ${\cal L}$ Lindbladian superoperator provides known co-efficients (the ME is fixed) to the quadratic monomials.  Note that we take the state vector to be unnormalised, with the normalisation included as an additional quadratic constraint.
The RHS of \erf{jumpCond} consists of cubic monomials arising from the state vectors (quadratic) multiplied by the transition rate parameters.  Assuming we are working with a minimally sized ensemble --- and a completely generic ME for which invariant subspace symmetries are not relevant --- the set of matrix equations (one for each ensemble member) forms a {\it polynomial} system containing  
\beq
\label{kmin}
K_{{\rm min}}D^2
\eeq
equations and as many (or slightly more) parameters. 
The lowest dimensional generic ME that has a PRE with $K_{{\rm min}}>D$ 
is for $D=3$, for which we expect $K_{{\rm min}}=5$ (when $L\geq 2$, allowing a fully connected PRE graph). Finding a PRE for such a case involves solving a square (it turns out there are no excess free parameters in this case) system of 45 polynomial constraints.  The simplest generic ME for which $K = (D-1)^2+1$ is not expected to be sufficient is for $D=3,L=1$, for which 
$K_{{\rm min}}=8$ (see \trf{table1}). Finding a PRE for such a case
involves solving a system with 72 polynomial constraints. 
In the case that the ME possesses an optimal invariant subspace symmetry, the polynomial system has fewer constraints:
\beq
\label{kminReal}
\frac{1}{2}K_{{\rm min}}(D^2+D).
\eeq
For the `re3it' ($D=3$, real valued state vectors), the invariant subspace that we choose for our detailed PRE search, $K_{{\rm min}}=4$ (provided $L\geq 2$) and the polynomial system has 24 constraints.  If $L=1, D=3$ then $K_{{\rm min}}=6$ (see \trf{table2}) and 36 constraint equations are obtained.  The question of whether it is feasible to solve such large systems, in order to find PREs, is now considered in some detail.

The most conceptually straightforward approach to solving such systems is via progressively eliminating variables, for example by calculating a {\it lex} Gr{\"o}bner basis~\cite{cox2006using} (a discussion of Gr{\"o}bner bases in the context of PREs can also be found in an appendix of~\cite{karasik2011tracking}).  However, even if the Gr{\"o}bner basis can be found, the elimination process typically rapidly increases the degree of the remaining monomials.  Unfortunately there are no general formulas for solving polynomial equations in a single variable of degree $\geq 5$ in a way that expresses the roots of that polynomial in terms of radicals.  This removes the possibility of finding PREs algebraically in these cases, so a numeric approach is essential.  Even numerically, it is still a very difficult problem and, in fact, it falls into the NP-complete complexity class~\cite{courtois2000efficient}.  It is important to realize that this difficulty extends to the decision problem of determining whether a solution exists~\cite{blum1996complexity}.  This is relevant as we will often be satisfied with finding an {\it example} PRE, and not all, PREs that exist for a given ME.  To illustrate, finding an example PRE for a specific ME with $K_{{\rm min}}>D$ and also proving that no PREs are possible for $K<K_{{\rm min}}$ (for the same ME) would provide evidence for Q3 and an answer to Q1.

Despite the computational complexity, polynomial systems {\it can} be numerically solved up to a moderate size.  Where is the boundary that separates the tractable from intractable?  Obviously it depends on the details of the polynomials and the applied solution methods, but for the purposes of this discussion the reader is provided with some circumstantial evidence from the literature as well as our direct experience.  In \cite{courtois2000efficient} it is suggested that even when using the highly efficient Gr{\"o}bner basis Faug{\`e}re F4 algorithm~\cite{faugere1999new} (a variant on the Buchberger algorithm~\cite{buchberger1976theoretical}) quadratic systems of size larger than 15 equations are very difficult.  The constraints of \erf{jumpCond} are cubic (except for $K_{{\rm min}}$ quadratic normalisation constraints) and are resultantly more difficult.  Our experience with Gr{\"o}bner basis techniques using MAGMA was that an example system of 16 equations was soluble in about 13hrs~\cite{privateSteel} whereas size 20 systems were out of reach (limitations of 200Gb of Ram were exceeded after several days runtime).  

The above discussion relates to cases in which we expect a solution to exist, rather than over-determined systems (number of variables less than number of constraints), for which no solution is expected.  In the latter circumstance, there is a remarkably useful result due to Hilbert that can allow a proof (analytically, via the obtaining of a Hilbert Nullstellensatz certificate of infeasibility) that no PREs of a particular size are possible.  In \arf{nullenstantz} we provide some more technical details of the procedure that we briefly outline here.  The algorithm~\cite{cox1992ideals} is as follows: \erf{jumpCond} provides a set of polynomial constraints that generate an ideal for which we calculate a Gr{\"o}bner basis.  If the basis contains $\{1\}$ then the polynomials have no common zero in $\mathbb{C}^{n}$ ($n$ being the number of parameters).  Unfortunately it is not straightforward to computationally extend this to a statement about common zeroes in $\mathbb{R}^{n}$ as the real numbers are not an algebraically closed field.  Thus, our technique will be to rule out complex solutions so that real solutions are also infeasible.  Despite this being heavy handed, it will allow several results to be established regarding PRE non-existence.  

Although obtaining a Hilbert Nullstellensatz certificate of infeasibility possesses the highlighted difficulty --- of finding a Gr{\"o}bner basis --- there are two reasons why we find the task simpler for ruling out PREs.  Firstly, it is simply the fact that the PREs that we wish to rule out are typically smaller than $K_{{\rm min}}$, so a smaller polynomial system is relevant.  Secondly, even for large systems, the presence of $\{1\}$ in the Gr{\"o}bner basis can quickly collapse the computation; we find that Gr{\"o}bner bases for such over-determined systems are typically easier to obtain.  The possibility of proving by these methods that there exist no PREs of a certain size positions us to directly tackle research questions Q1-Q2, and provide evidence for Q3a.

Coming back to the difficulty of actually finding solutions to polynomial systems, when they exist, we were clearly motivated to explore different (beyond Gr{\"o}bner basis) numerical techniques.  Further success was had with the method of polynomial homotopy continuation, in which the problem is cast into an `embarrassingly parallel'~\cite{verschelde1999algorithm} form.  As a brief explanation, solutions of a simple `start system' (a polynomial system that possesses the desired features) are tracked (see endnote~\cite{trackingComment}) as the system is homotopically continued (transformed) to the target polynomial system, with each solution path able to be treated independently (a more in-depth discussion is provided in \arf{PHCpackMonodromy}).  This is particularly suited to our goal of finding example PREs as we can stop tracking paths once a single PRE solution is found.  That is, the entire search space does not have to be be explored.  This is by no means a panacea as the number of paths can be enormous and a large number may have to be followed before either a solution is found or it is decided that resources are better spent exploring a different ME.  For example, the B{\'e}zout bound for the maximum number of solutions to a system is the product of the largest degree of the polynomial equations; admittedly this is typically much larger than than the tightest available bounds for sparse systems containing relatively few monomials compared to parameters.  The B{\'e}zout bound for $K=5$, $D=3$ system evaluates to $3.9\times 10^{20}$ potential solutions and that for the $K=8$, $D=3$ system is $8.8\times 10^{32}$.  A particularly useful extension to the polynomial continuation approach is the very recently developed Monodromy method~\cite{duff2016solving}, which reduces the number of paths that have to be tracked and, importantly, eliminates the necessity of doing a costly pre-computation to obtain a bound tighter than that of B{\'e}zout.  This is also described further in \arf{PHCpackMonodromy}.  

Utilising the polynomial homotopy continuation numerical approach as well as the monodromy extension, PREs were extracted from systems of 24 polynomial equations.  Despite this being less than that required to investigate generic $K=5, D=3$ PREs, it {\it is} sufficient to find PREs that possess the invariant subspace symmetry, for which $K_{{\rm min}}=4$ for $D=3$.  The result is that, by combining symmetry considerations and newly applied numeric methods, we are now capable of investigating research question Q3b described in the introduction.

\section{Results}
\label{results}
In this section the previously developed counting arguments and symmetry techniques are used to guide the finding of PREs or to rule out their existence for a large number of example MEs.  The examples are chosen from classes (specified by $D,L$) that allow us to answer the following two existential questions raised in the introduction (following~\cite{karasik2011many}): {\it are there MEs for which the minimally sized PRE is larger than $D$?} (Q1) and {\it is an ensemble size of $K=(D-1)^2+1$ always sufficient for a PRE to be found?} (Q2).  Evidence is also accumulated as to the validity of parameter and constraint counting being used to rule out PREs (Q3a) or determine that they are feasible (Q3b).  The existence of an explicit measurement scheme that realizes each identified PRE is confirmed to ensure that the PRE is not merely a mathematical construction.  

When we prove that no PRE can exist, the proof consists of a Hilbert Nullstellensatz certificate of infeasibility (see \srf{computationalDifficulty} and \arf{nullenstantz}).  To numerically search for example PREs in cases where they are expected to exist, we use the technique of polynomial homotopy continuation (see \srf{computationalDifficulty} and \arf{PHCpackMonodromy}).  The software that we used was PHCpack~\cite{verschelde1999algorithm} together with a recently developed monodromy package~\cite{duff2016solving} that extends its capability.  

To aid the reader, a summary of our results is provided in~\trf{systemSummary}, to which we will refer. 

\begin{table}
 \centering
\begin{tabular}{cccccccc}
 \centering
\textbf{Class}&{\textbf{ \specialcell[b]{Symm.\\ Type}}}&
 \textbf{K} & \textbf{L} &\textbf{Graph} & \textbf{\specialcell[b]{PRE\\ exists?}} & \textbf{Method}&
% \textbf{\specialcell[b]{Timing\\ (secs)}} &
 \textbf{Purpose} \\ \hline
%I&N&3&1&FC&N&MAGMA&8\\
I&N&3&2,3&FC&N&Gr{\"o}bner&Q1\&Q3a\\
%II&N&3&1&FC&N&Gr{\"o}bner&8&Q1\\
II&N&3,4,5&1&R&N&Gr{\"o}bner&Q1\&Q3a,Q1\&Q3a,Q2\&Q3a\\
III&S&3,4&3&FC&N,Y&Gr{\"o}bner,PHC&Q1\&Q3a,Q3b\\
IV&SW&3,4&2&FC&N,Y&Gr{\"o}bner,PHC&Q1\&Q3a,Q3b
\end{tabular}
\caption{Classes of $D=3$ MEs for which we obtained major results are labeled and described.  The second column describes the symmetry (there can be more than one) of the ME: either (N)one, invariant (S)ubspace symmetry (\srf{invSubSec}) or (W)igner symmetry (\srf{wigSymmSec}).  The third column gives the considered ensemble size, while the fourth column states the number of decoherence channels.  The `Graph' column states whether the graph of the ensemble was fully connected (FC) or was necessarily (R)estricted by the rate counting arguments of \srf{rateCounting}.  The sixth column indicates what type of PRE existence proof was obtained: N indicates that a Nullstellensatz proved that no PRE exists, while Y indicates that a PRE exists, with the proof made by example.  The `Method' column indicates how the computation was performed: either via Gr{\"o}bner basis with MAGMA or polynomial homotopy continuation (PHCpack and monodromy extension).  The final column indicates the significance of the result by referring the to the research question that it addressed.  Comma separated values in columns $3$ or $4$ (never both) indicate multiple investigations.  In subsequent columns, if different values apply respectively to each investigation then these are also indicated by comma separated values.}
\label{systemSummary}
\end{table}

\subsection{Are open quantum systems harder to track than open classical systems?}
\label{majorQ}

In order to answer this central question in the affirmative, it needs to be proven that there exists a ME such that $K=D$ states are not sufficient for a PRE to be formed (Q1).  This is because a classical system can always be tracked with $K=D$ states as the occupation ($1$ or $0$) of each state could, in principle, always be known by monitoring the environment.  It is also of interest to look for a generic difference in difficulty of tracking quantum and classical systems. That is, we ask whether $K=D$ is insufficient {\it in general}, for randomly drawn MEs.  By addressing this, in a manner guided by our heuristic, we will also gather evidence regarding to Q3a.

For the moment, we delay the task of actually finding a PRE and concentrate on disproving their existence for $K<K_{{\rm min}}$, with $K_{{\rm min}}$ determined utilising our ($L$ and symmetry dependent) knowledge of parameter and constraint counting.  In $D=2$ it was shown in \crf{karasik2011many} that $K=D=2$ is always sufficient to find a PRE, so the search for an example system for which $K$ is necessarily larger than $D$ is extended to $D=3$.  (All the results we obtain in this paper for specific example MEs apply to $D=3$.)  In \srf{countArguments} we found that for a {\it generic} ME (no symmetry considerations), described by a sufficiently large number of linearly independent decoherence channels ($L\geq 2$ for $D=3$), the number of parameters equals (or exceeds) the number of constraints for a minimum sized ensemble $K_{{\rm min}}=5$ (a summary for $K_{{\rm min}}$ in terms of $L,D$ was provided in \trf{table1}).  Thus, it is logical to attempt to rule out PREs by Hilbert Nullstellensatz for $K=3,4$.  For completely generic $D=3$ MEs (which  we label as class I, see \trf{systemSummary}), it {\it is} within our computational ability to obtain the $K=3$ certificates of infeasibility, but, unfortunately, $K=4$ proved too difficult.  However, it should be remembered that our current goal is only to find example MEs that require $K>D$ ensemble members to form a PRE, so ruling out $K=3$ is sufficient in this respect.

To select generic MEs, the $D=3$ Hamiltonian and Lindblad terms were parameterised according to
\bqa
\hat{H}=i\left(
    \begin{array}{ccc}
  \frac{1}{2}\left( \alpha_1-\alpha_{1}^{*}\right)&\alpha_2&\alpha_3   \\
  -\alpha_{2}^{*} &\frac{1}{2}\left( \alpha_4-\alpha_{4}^{*}\right)&\alpha_5\\
  - \alpha_{3}^{*} &-\alpha_{5}^{*}&\frac{1}{2}\left( \alpha_6-\alpha_{6}^{*}\right)
    \end{array}
    \right), \quad \hat{c}_{l}=\left(
    \begin{array}{ccc}
  \gamma^{l}_1& \gamma^{l}_2& \gamma^{l}_3   \\
    \gamma^{l}_4 & \gamma^{l}_5& \gamma^{l}_6\\
    \gamma^{l}_7 & \gamma^{l}_8& -\gamma^{l}_1-\gamma^{l}_5
    \end{array}
    \right),
\label{meParams}
\eqa 
with $l={1,...,L}$.  Note that the Lindblad's are taken as traceless without loss of generality~\cite{doi:10.1063/1.522979}.  The complex values $\vec{\alpha},\vec{\gamma}^{l}$ were chosen randomly from a uniform distribution covering some rectangular range $[-a-ia,a+ia]$, with $a$ being some arbitrarily chosen cutoff (we took $a=3$). That is, $6+8L$ complex random values were generated in order to specify an example ME from class I, with $L$ chosen sufficiently large (for convenience we focused on $L=2,3$), so that there was no rate counting restriction on the graph being fully connected.  Once $\hat{H}$ and $\hat{c}_{l}$ have been specified, the constraints that must be satisfied by a potential PRE were formed.  
These are found in \erf{jumpCond} (see \erf{me1} for the Lindbladian definition), with 3 sets  (remembering that we are investigating $K=3$) of 9 equations (27 total) compared with 21 state vector and transition rate parameters.  These polynomial systems were fed into MAGMA for 20 different example class I MEs, with a Hilbert Nullstellensatz achieved for each system.  It is curious that there was a large range of difficulty in obtaining the computational proofs: around half completed in about 8 seconds but the rest took from hours up to $2.5$ days in the hardest case.  Having found MEs for which $K=D$ ensemble members are not sufficient to form a PRE (and therefore answered Q1), we then state that open quantum systems {\it are} harder to track than open classical systems.  Indeed, we believe them to be generically so.  That is, we conjecture that the proportion of generically selected class I MEs that possess $K=3$ PREs will be vanishingly small.  That the heuristic predicted no $K=3$ PREs would exist, for the selected MEs, is evidence for its accuracy (and addresses Q3a).  To ensure that our results can be reproduced, a specific example of every class of ME that we obtain a result for is given in \arf{repro}.

\subsubsection{$L=1$ MEs are harder to track}

We have shown that open quantum systems are harder to track than open classical systems, but only in a minimal sense.  That is, we proved that $K=D=3$ PREs do not exist for randomly generated MEs without providing evidence for how large $K$ must be to find a PRE.
As mentioned earlier, $K=4,D=3$ Nullstellens{\"a}tze for generic MEs are currently beyond us, but we can make progress for the particular case of $L=1$.  This is because the number of parameters is limited in this instance by our rate-counting arguments, resulting in a more highly over-determined system of constraints.  This conspires to make a computational Nullstellensatz less demanding to achieve in the systems we investigated.  In the case when $L\geq 2$, there are 36 constraint equations and 32 parameters for a $K=4$ PRE, but when $L=1$ the number of parameters reduces to 26.  This is despite the optimal `rate packing' graph being chosen, see \srf{rateCounting}.  If the graph is non-optimal, then fewer than 26 parameters are present.  
This allowed us to obtain $K=4$ Nullstellens{\"a}tze for all 30 example MEs that we chose according to the method above (but now with $L=1$).  To distinguish this result we define the generic $L=1$ MEs as belonging to class II.  

Somewhat surprisingly, the same technique was also successful in ruling out $K=5$ PREs in all 30 class II examples.  In this case, there are 45 constraints with a maximum of 36 parameters.  It is perhaps the relatively large ratio of constraints to parameters that allows us to obtain a Nullstellensatz in this case, while the 36 constraint and 32 parameter scenario for generic $K=4,L\geq 2$ MEs was out of reach.  
 We provide a brief discussion of further difficulties of this $L=1$ computational proof in endnote~\cite{subOptimalRates}. 
To be clear: there is no surprise surrounding the non-existence of $K=5$ PREs for $L=1$ as they are not expected according to \trf{table1} ($K_{{\rm min}}=8$).  
Rather, the surprise is that we were able to prove it via Nullstellensatz.  The implication of the $L=1$ result is that we have proved, for some example MEs, that $K=(D-1)^2+1$ states are {\it not} sufficient to form a PRE.  This answers an open question of \crf{karasik2011many} that was raised as Q2 in the introduction.  We leave the task of attempting to obtain Nullstellens{\"a}tze for $K=6,7$ PREs belonging to generic $L=1$ MEs for future work.  

Note that if a state {\it matrix} parameterization of ensemble members had been used, as discussed in~\srf{dimCounting}, then it would be concluded that $K=5$ states are sufficient for the number of parameters to exceed constraints in a class II ME.  The fact that a $K=5$ Nullstellensatz is obtained for specific MEs is then evidence supporting the state vector parameterization, which gives a minimal parameterization of pure states.  
That is, as is worth emphasizing, the results of this subsection are completely consistent with our parameter and constraint counting arguments --- we have been able to rule out PREs in many cases when our heuristics would not expect them to exist, thus providing strong evidence for an affirmative answer to Q3a. In particular, our heuristics say that $L=1$ MEs are generically harder to track than would have been expected in the absence of consideration of $L$, 
and our numerics strongly support that conclusion.

\subsection{Applying symmetry in order to find PREs}
\label{invSymmResults}
In the previous subsection we showed that open quantum systems are harder to track than open classical systems, with the minimal $K=D=3$ PREs ruled out generically and $K\leq5$ PREs typically ruled out for $L=1$.  Notably, we have not as yet actually found any PREs!  To do so would place an upper bound on how difficult example MEs are to track and potentially provide confidence in our heuristic arguments as to when PREs are expected (which forms Q3b).  For $D=3$,  PREs are only expected to exist for generic class I MEs (with $L\geq2$) when $K\geq 5$.  However, as discussed in \srf{computationalDifficulty}, this leads to polynomial systems of at least 45 equations and 45 parameters that are very difficult to solve, even numerically. In this subsection we do find PREs, but only after the application of symmetry makes the task tractable. 

\subsubsection{Invariant subspace symmetry}
The first strategy we use is to introduce the invariant subspace, $\mathfrak{D}_\mathfrak{I}$, of \srf{invSubSec} in which the image of $\mathfrak{D}_\mathfrak{I}$ under $e^{{\cal L}t}$ for any $t\geq 0$ is $\subseteq \mathfrak{D}_\mathfrak{I}$. When we restrict the PRE to
$\mathfrak{D}_\mathfrak{I}=\R^{D\times D}\cap\mathfrak{D}\left({ \mathbb{H}}\right)$, it was shown that $K_{{\rm min}}$ is reduced to 4 for $D=3$ and we are led to a set of 24 constraints in 24 parameters.  The simplest way to achieve this is to choose the Hamiltonian and Lindblad terms of the same form as \erf{meParams} but with $\vec{\alpha},\vec{\gamma}^{l}$ now consisting of purely {\it real} random parameters (in contrast to the previous complex-valued choice).  For simplicity, we restricted to $L=1,2,3$. 
We created 240 different MEs by sampling each parameter of $\vec{\alpha},\vec{\gamma}^{l}$ 240 times over the uniform distribution $[-3,3]$.  These MEs formed class III of \trf{systemSummary}.  

The randomly sampled $\hat{H},\hat{c}_{l}$ (purely imaginary and real, respectively) were then used to form the constraints governing PREs, \erf{jumpCond}, with an ensemble size of $K=4$.  To ensure that the imaginary constraints were automatically satisfied, the PREs were parameterised as being real-valued (that is, as re3its).  Despite having an equal number of parameters and constraints, a PRE is not guaranteed to exist.  In our findings, this was manifest as PREs were only found for a fraction of the tested MEs.  It is worth mentioning that not all the solution space was searched for each ME as the goal was to find solutions, not rule them out by numerical means (we use Nullstellens{\"a}tze for non-existence proofs when $K<K_{{\rm min}}$).  The numerical method we used was polynomial homotopy continuation~\cite{verschelde1999algorithm} (see \arf{PHCpackMonodromy} for details).  Typically we would run a multi-hour search with parallel processing for each class III ME, with multiple systems run also in parallel.  Of the 240 example MEs that had the re3it as an invariant subspace, one third were chosen with $L=3$, one third with $L=2$ and the final third with $L=1$.  A more thorough investigation for higher $L$ is left to future work.  Of the 80 MEs with $L=3$, PREs were found for 6 ($7.5\%$).  Interestingly, zero PREs were found for the $L=2$ sample, indicating that the occurrence rate of PREs in this ME class is likely to be very low.  It might even be of concern that $L=2$ PREs were ruled out by some undetermined cause, throwing into doubt our rate counting argument.  This, however, is refuted once we apply the further unitary symmetry in the 
following subsubsection, where we do find $L=2$ PREs.  The rationale behind $L=2$ PREs being less generically prevalent is that there is less freedom for the experimentalist in selecting an appropriate adaptive measurement scheme.  In \erfs{jumpOps}{noJumpOps} the ${\bm S}$ matrix (the dimension of which scales with $L$) was introduced which represents mixing of the system output fields.  For larger $L$ there are more free parameters that the experimentalist may vary.  This is despite the same number of parameters and constraints describing the PRE in \erf{jumpCond}.  One way of thinking about it is that decreasing $L$ causes step-changes in $K_{{\rm min}}$ but that this does not capture the entire effect --- the difficulty of satisfying constraints when $K_{{\rm min}}$ is static is increasing also.  We have not given a quantitative theory of this and it is something that could be investigated in the future.  As the reader will anticipate, no $L=1$ PREs were found.  They are not expected as in this case $K_{{\rm min}}=6$, as can be seen in \trf{table2}.  In fact, they were proven to be impossible for some MEs when $K<K_{{\rm min}}$ in the previous subsection.

By finding PREs for $L=3$ MEs with the $K=4$ ensemble size expected (symmetry considerations reduced $K_{{\rm min}}$ to 4), we have provided evidence that our heuristic for finding PREs is accurate.  It has also been accurate in ruling them out for $K<K_{{\rm min}}$, as verified by Nullstellensatz; this result is consistent with \srf{majorQ} (where MEs not possessing the invariant subspace symmetry were considered).  %\pw{dont think its a really new result}
%; we made sure to verify this, by Nullstellensatz, for the MEs that we obtained PREs for.
%(it has also been accurate in ruling them out for $K<K_{{\rm min}}$ --- we made sure to perform this by Nullstellensatz for the MEs that we obtained PREs for).  
Another compelling demonstration of the heuristic is given by considering the perturbing of the ME in such a way that it causes a step-change in our perceived likelihood of a PRE existing.  The way to implement this is to take the ME possessing an invariant subspace for which we have a PRE and then perturb it.  First we perturb it within the invariant subspace.  For very small ME changes the PRE can be `tracked' using a numerical optimization in which the sum of the squares of the constraints are minimized.  As the new PRE (if it exists) will almost certainly lie very close to the old PRE, the latter can be used as an initial seed and the new PRE found.  This is a crude form of a `cheaters homotopy'~\cite{li1989cheater}, where a polynomial system with the same co-efficient structure is repeatedly solved.  Second, we perturb the ME in a way that breaks its symmetry (for example, a small imaginary component can be added to a few of the Lindblad parameters).  Then the imaginary constraints of \erf{jumpCond} can no longer be ignored.  This returns the polynomial system to being over-determined and no solution is expected.  This was investigated for a number of MEs and solutions were only found for PREs when the ME was perturbed in a way that maintained its invariant subspace symmetry.  To summarize, extremely small ME changes impact in a stepwise fashion the existence of PREs in exactly the manner forecast by our heuristic, thus answering affirmatively Q3b (and providing more evidence for Q3a).

\subsubsection{Wigner transformation group symmetry}

\label{uniSymmPREs}
One of the exciting results of \cite{warwis2018a} was that the polynomial system size required to find a PRE can be dramatically reduced when the Wigner (unitary/antiunitary) symmetry of \erf{Uinv} is imposed upon both the ME and the PRE (as per \erfs{induced}{inducedMap}).  We then found that the constraints of \erf{jumpCond} that are applicable to ensemble members within the equivalence class of \erf{equivClass} are redundant in the sense that if they are satisfied for one member then they hold for all members.  The consequence is that example PREs for MEs possessing the symmetry of \erf{Uinv} are much easier to find.  The reader is reminded that not all PREs pertaining to a ME can be found in this way as there could be PREs not possessing the symmetry despite the ME being symmetrical.  Given the great reduction in computational complexity, it should be possible to find PREs for larger systems than we consider in this paper.  For example, if a $D=4$ ME having the re4it as an invariant subspace (leading to $K_{{\rm min}}=7$, see \trf{table2}) has, in addition, a $\mathbb{Z}_3$ {\it unitary} symmetry then the graph nodes representing the ensemble could break into two triplets and a singlet.  Counting non-redundant constraints gives 35, which is larger than we have solved to this point but potentially within our short-term future capabilities.

In this paper we content ourselves with performing the important task of verifying that Wigner symmetry can be successfully applied to find new PREs in $D>2$.  To this end, a $D=3$ system with a unitary $\mathbb{Z}_{2}$ symmetry {\it and} a re3it invariant subspace  was considered.  These form MEs in class IV.  As we have seen, $K_{{\rm min}}=4$ for this system.  The $\mathbb{Z}_{2}$ symmetry allows the 4 ensemble members to break into two doublets, with the result being that the square polynomial system of size 24 breaks into two, square, size 12 systems.  The systems being square means that the results of this subsection relate to Q3b.  The number of constraints is small enough that they can be solved completely by Gr{\"o}bner basis methods (and, of course, the polynomial homotopy continuation method) to find PREs, when they exist.  Unlike the larger systems, for which only a portion of the symmetric solution search space is covered, when no PREs are found then we can be `sure' (technically only up to the utilized numerical accuracy~\cite{numericalCertainty}) that none exist possessing the unitary symmetry.  (As always, we stress that there may be PREs not possessing the symmetry). The fact, detailed below, that some unitarily symmetric MEs lack PREs possessing the unitary symmetry gives us reason believe that having a sufficient number of parameters relative to constraints is generically a necessary, but not sufficient, condition (that is, a heuristic only) for PRE existence.

%not a necessary and sufficient
To provide a concrete example, let us choose the form 
$$\hat{H}=i\alpha\left(\ket{1} \bra{3}-\ket{3} \bra{1}\right), 
\ L_{1}=\gamma^{1}\left(\ket{1} \bra{2}+\ket{2} \bra{3}\right), \ L_{2}=\gamma^{2}\ket{3} \bra{1}, \ L_{3}=\gamma^{3}\ket{2} \bra{3},$$ which possesses the \erf{Uinv} unitary symmetry with ${\cal T}\ketbra{\psi}{\psi}\equiv\hat{U}\ketbra{\psi}{\psi}\hat{U}^{\dag}$ and $\hat{U}=\mathbbm{1}-2\ket{2} \bra{2}$ and also has a re3it invariant subspace, as long as $\alpha,\gamma^{1},\gamma^{2},\gamma^{3}$ are real.  To illustrate how the $\mathbb{Z}_{2}$ symmetry breaks the PRE into pairs for this particular ME we provide a schematic in \frf{Z2PRE}.  
Note that this unitary $\mathbb{Z}_{2}$ symmetry sends $\ket{2}\rightarrow -\ket{2}$.  As we were able to find PREs for a finite fraction ($3\%$ out of a sample size of 200\blk) of randomly chosen MEs of the described form, we are confident that Wigner transformation symmetry will be an important tool going forward for the study of larger PREs.  Interestingly, we were also able to find $L=2$ PREs (achieved here by setting $\gamma^3=0$), which is in contrast to \srf{invSymmResults}, where only the re3it symmetry was applied.  This is consistent with the heuristic which states that PREs should be possible for $L=2$ in $D=3$.  In case it be thought that the $L=2$ PREs only exist for a set of measure zero MEs (possessing the $\mathbb{Z}_{2}$ symmetry), we break this symmetry by perturbing the ME and perform a cheaters homotopy to obtain the perturbed PRE. 
This shows the usefulness of symmetry as a method to obtain solutions even for systems where that symmetry is broken. 
 As the $L=2$ case is interesting for multiple reasons it is the explicit example we give, for class IV, in \arf{repro}.  \blk

\begin{figure} [t]
\centering
\includegraphics[scale=0.6]{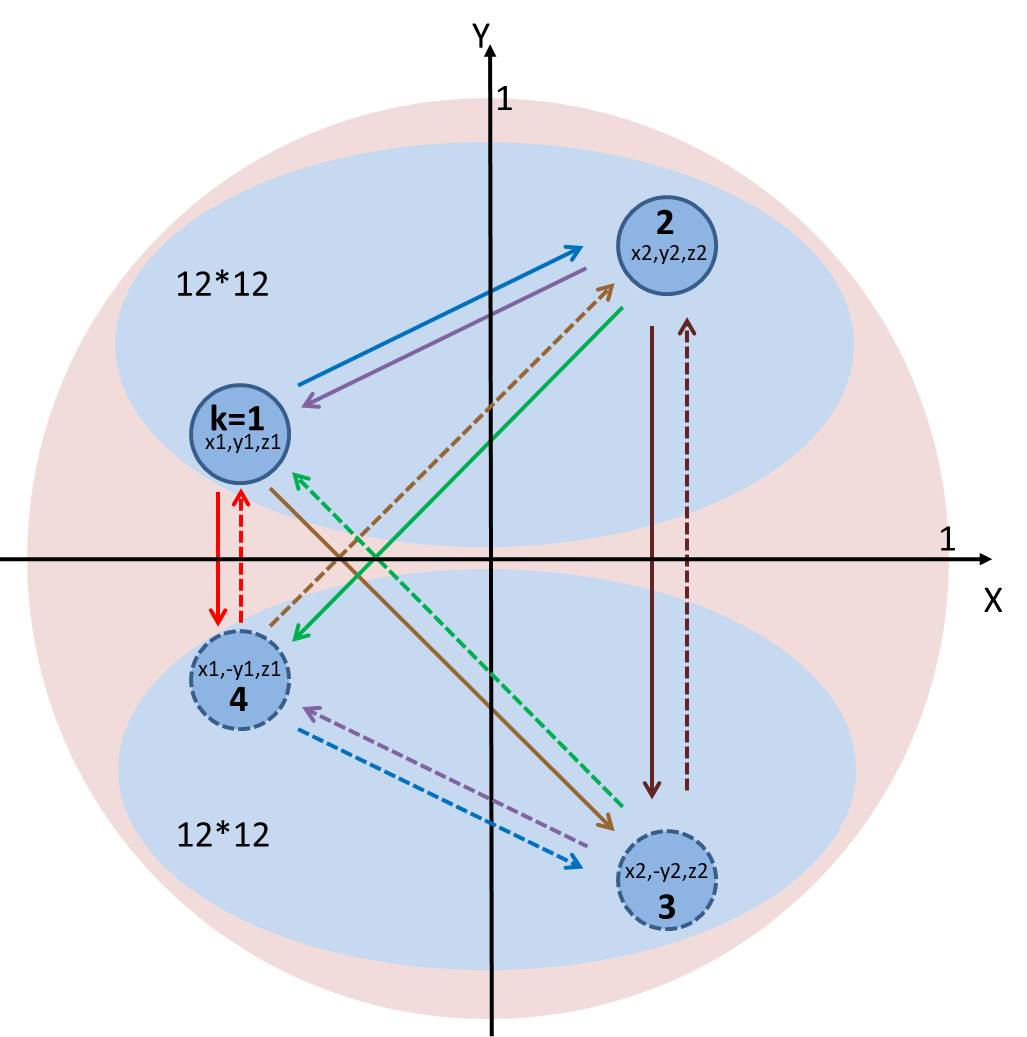} 
\vspace{-2mm}
\caption{A hypothetical PRE is depicted for a ME having a $\mathbb{Z}_{2}$ Wigner symmetry described by $\hat{U}=\mathbbm{1}-2\ket{2}\bra{2}$.  The $K=4$ ensemble breaks into 2 pairs (indicated by the blue ovals) that are related by the unitary symmetry, as do the set of constraints describing them.  Thus a 24 equation, 24 variable system of polynomials ($24*24$) splits into two $12*12$ systems, with the solution of the second system fixed by that of the first. The values $x,y,z$ represent the re3it via $\ket{\phi_{k}}=x\ket{1}+y\ket{2}+z\ket{3}$ with $x^2+y^2+z^2=1$.  Thus, the $z$ value (assumed $>0$ to fix the overall phase) can be inferred from the position of the node (state) on the unit disk (the pink region).  Everything dashed in the figure is obtained by reflection about the $x$-axis including the transition rates.  
 }
%\vspace{-4mm}
\label{Z2PRE}
\end{figure}

\subsubsection{Measurement schemes}
It is interesting that one can find a PRE, via solution of \erf{jumpCond}, but not immediately know how to realize it in the laboratory.  That is, the adaptive measurement scheme parameters ${\bm S}^{k},\vec{\beta}^{k}$ are still to be found.  The method of doing so is decribed in \arf{finMeasScheme} and, not surprisingly, involves solving polynomial systems, albeit of a typically much easier nature than those of the PREs themselves.  We were able to find the explicit measurement scheme to realize each of the PREs found during our investigations.  Additionally, schemes were found that had the expected Wigner symmetry and also possessed the invariant subspace symmetry, in that they were real-valued (the reader is referred to the appendices of \cite{warwis2018a} for a discussion of measurement scheme symmetry, in particular Wigner symmetry).  It is worth noting that in all cases multiple schemes were found to realize the same PRE; for each PRE we were able to find a scheme that alternatively did and did {\it not} have the symmetry possessed by the ME.  This highlights how, analogously, PREs that break the ME symmetry can be expected, despite their being harder to find.  That measurement schemes were found to realize our PREs is consistent with \cite{karasik2011many} where the necessary schemes were also obtained. However, there, the analysis beyond $K=2$ was restricted to cyclic jumps for PREs of size $K=3$ (still within $D=2$), allowing an analytic form of the measurement scheme to be found given the PRE. Here we find many examples that are not cyclic, the significance of which is further discussed in \srf{Q3}.  An example measurement scheme that realizes a $D=3$ PRE is given in \arf{repro}.

\section{Conclusion}
\label{conc}

\subsection{Summary}  

In this paper, we have considered the question of whether tracking an open quantum system is harder than tracking an open classical system of the same size, $D$. Here, by `tracking', we mean undertaking a continuous measurement on the environment of the system in such a way as to obtain enough information so that the conditioned state of the system is pure, but without disturbing the average evolution of the system. This average evolution is assumed to be described by a Markovian master equation (ME), in either the quantum or classical case, and to possess a unique steady state having rank $D$.
We are quantifying the difficulty of tracking a system by the the minimum number of macroscopic states $K$ that the  classical measurement device requires. To put it another way, $K$ is the minimum number of distinct pure states which can make up a physically realisable ensemble (PRE), the collection of the conditioned system states at all times. 
Obviously in the classical case, $K=D$ is sufficient and necessary for tracking.  In order to structure our investigation we framed our research around the three major open questions, given in \cite{karasik2011many,karasik2011tracking}, that were presented in our introduction; we now summarize our answers to these questions.

\subsubsection{Q1: {\it Are there MEs for which the minimally sized PRE is larger than $D$?}} \label{Q1}
We have answered this question in the affirmative and hence shown that open quantum systems can be harder to track than open classical systems.  Our investigation was carried out in $D=3$ ---  this is the smallest dimension of interest (it is known that in $D=2$ that $K=2$ PREs always exist~\cite{karasik2011many}) --- for a random selection of 240 MEs.  For each of these MEs, we obtained a computational proof, in the form of a Hilbert Nullstellensatz certificate of infeasibility for the equations governing PREs, that $K=D$ PREs cannot exist.  For $D\geq 3$ we expect this to hold generically, so that only an infinitesimal fraction of MEs will have $K=D$ PREs.  Our result was not unexpected; the parameter and constraint counting arguments of \crf{karasik2011many} suggest that $K\geq (D-1)^2+1$.  Whether this is strictly an equality then formed our next question.

\subsubsection{Q2: {\it  is an ensemble size of $K=(D-1)^2+1$ always sufficient for a PRE to be found?}} \label{Q2}
The answer to this is `no', as we found example MEs which provably (via Hilbert's Nullstellensatz) have no PREs of size $K\leq (D-1)^2+1$. The search for such examples was guided by the development of a new parameter counting argument that provided a heuristic as to when there is a significant possibility of a PRE existing.  This heuristic, which is restated for clarity in \srf{discuss}, differs from that of \crf{karasik2011many} as it has a dependence upon the number of decoherence channels, $L$, of the ME.  This allowed us to form an expectation that generic, randomly chosen, $L=1$ MEs in $D=3$ would not possess PREs with $K< 8$. This was subsequently partially confirmed as we proved that $K\leq(D-1)^2+1=5$ PREs were impossible for some randomly chosen $D=3, L=1$ MEs. Further assessment of the newly formed heuristic was our next research question.

\subsubsection{Q3: {\it  does the refined parameter counting heuristic reliably predict whether PREs are feasible for a ME of a given form?}}\label{Q3}

Our studies suggest a positive answer to this question.  However, we cannot answer `yes' to Q3 with the same absolute confidence as we answered the preceding two questions. The reason for this is the multi-faceted and open-ended nature of Q3, as we now detail. 

We first consider whether the heuristic's prediction of ruling out PREs for ensembles smaller than the determined threshold is accurate (this is Q3a in the introduction). In this initial regard we note that, whenever the heuristic could be tested (limited by finite computational resources), it provably made the correct prediction.  These tests consisted of MEs for a $D=3$ system 
with PRE size $K=3$ for $L=2,3$ (see Q1) and $K\leq 5$ for $L=1$ (see Q2).  A notable topic for future work would be to extend the Nullstellensatz results for $D=3$, $L=1$ to $K= 6>(D-1)^2+1$, as this would allow us to establish that the prior heuristic, of \crf{karasik2011many}, provides poor guidance even when $K$ exceeds its minimal requirement.  This is a stronger result than that established in \srf{Q2}. 
\blk

If, as it seems, the refined heuristic is correct in predicting the non-existence of PREs, then it follows that 
for generic quantum MEs, regardless of $L$, the minimum PRE size equals $D^2$ to leading order in $D$. This is quadratically larger than the $K=D$ result that pertains to classical system. That is, open quantum systems are not only harder to track than classical systems (see Q1); 
they are, it appears, harder to track by a factor that increases without bound as the dimensionality increases. 

The accuracy of the developed heuristic's guiding of when PREs {\it are} expected, was posed as Q3b. Evidence for this accuracy was obtained by finding actual PREs for a finite fraction of randomly drawn MEs with the ensemble size that was predicted to be sufficient. To obtain this evidence, however, required the introduction of symmetry to the ME and PRE.  An invariant subspace ME symmetry was considered in which the PRE was assumed to live in the subspace (we chose the Lindbladian to preserve the re3it subspace).  This reduced the effective size of the polynomial system that had to be solved, with the predicted minimum ensemble size also reduced to $K_{{\rm min}}\blk =4$.  A moderate sized set of randomly chosen MEs (that obeyed the symmetry) were examined using the numerical method of polynomial homotopy continuation.  This led to the finding of PREs for 6 of them, which was 7.5\% of the sample size for those MEs that had $L=D=3$.  By introducing a further $\mathbb{Z}_{2}$ unitary symmetry, MEs that had PREs for $L=2$ were also found with $K=4$, although in this case only $3\%$ out of a sample of 200.  As expected, none were found for $L=1$, where $K_{{\rm min}}=6$.  Note that explicit measurement schemes able to realize each of the discovered PREs were identified.

As a final discussion point regarding Q3, we revisit the possibility, raised in the introduction as Q3c, of investigating PREs for MEs and ensembles that lead to a larger number of parameters than constraints.  Intuitively, the additional parameters make the algebraic requirements easier to satisfy and PREs are expected to exist for a larger fraction of MEs than when the ensemble size is minimal. Additionally, one might expect that when there are more parameters than constraints, there exist MEs for which the number of PREs of a particular size becomes infinite, rather than there being merely isolated solutions.  In fact, preliminary investigation confirms this expectation --- for the case of $D=2$, $K=3$ with $L=1$, lines of PRE solutions on the Bloch sphere are found when an additional transition is allowed.  These were not 
found in \cite{karasik2011many} because that paper restricted to cyclic 
jumps; by not making that assumption we end up with more parameters than constraints. 

\subsection{Discussion and future work}
\label{discuss}
We begin our discussion by noting that Q3 was phrased in terms of the number of parameters and constraints, rather than just comparing $K$ with $K_{{\rm min}}$.  This choice was made as it is possible, due to the integer nature of $K$, to have $K=K_{{\rm min}}$ but for there be more parameters than constraints.  Despite this, it is worth re-stating the expression for $K_{{\rm min}}$, in the form applicable to generic MEs, in our concluding remarks as it represents a 
concise and clear facet of our extension of the work of \crf{karasik2011many}:
\beq
K_{{\rm min}}= (D-1)^{2}+1+{\bf 1}_{\{L< D-1\}}\times\left(2D-2L-1 \right).
\label{simpleKMin2}
\eeq
The term multiplying the indicator function enlarges the expected minimum PRE size, from that predicted by \crf{karasik2011many}, but only in the case that $L<D-1$. 

It is natural to wonder whether there exists other criteria, beyond $L$, that imply an increase in  the lower bound on $K_{{\rm min}}$.  That this may be the case is suggested, for example, by the fact~\cite{karasik2011many} that in $D=2$ it was shown that a small ratio of Hamiltonian to Lindblad operator magnitude led to more $K=2$ PREs existing for a particular ME.  Whether this is true in higher dimensions, and for generic, MEs is yet to be looked at, but the point that we wish to make here is that the various properties of ${\cal L}$ play a central role in defining the characteristics and existence of PREs.

Our work has established the usefulness of having knowledge of $K_{{\rm min}}$, but it is, as we have emphasized, merely a heuristic, and there is no guarantee that a PRE of size $K_{{\rm min}}$ will exist.  However, on the other hand, we have not proven, for any ME, that there is no $K_{{\rm min}}$-sized PRE: we have failed to find them in some cases, but only with an incomplete search.  (A Nullstellensatz proof of non-existence is not available when $K=K_{{\rm min}}$ as there will exist complex-valued solutions that are ineligible to be PREs.)  A future task of interest is, therefore, to conduct a complete search for some sample MEs that rules out a PRE of size $K=K_{{\rm min}}$.  A further strong hint that this is to be expected was obtained in \srf{invSymmResults} where symmetry was used to break the polynomial system down into a small enough size that the entire space could be searched, with a negative outcome.  However, in that case, generic PREs were not considered, merely those possessing the symmetry.

Despite providing answers to several open questions raised in \crf{karasik2011many}, it is notable that the portion of our results that are example-based concern low dimensional systems. We have gone beyond the qubits of \crf{karasik2011many}, but only to $D=3$. The computational difficulty of finding PREs is such that simplifying techniques are necessary to make tractable their detailed study.  Indeed, even in $D=3$ we found necessary the introduction of an invariant subspace and unitary symmetry --- this served to greatly reduce the size of the pertinent polynomial system.  The methods of applying unitary symmetries, or, more generally, Wigner transformation symmetries, that have been developed in \cite{warwis2018a} and utilized in this paper, provide an exciting opportunity to progress to higher dimensional systems of interest.  For example, the study of a $D=4$ composite system (two qubits) in terms of the statistics and dynamics of entanglement of discovered PREs would be intriguing, and is a topic for future work.

Of course, there are many, PRE related, computationally difficult tasks for which the application of symmetry is either not appropriate or does not fully alleviate the problem.  An example of this is obtaining proof that $K<K_{{\rm min}}$ PREs are ruled out for a sample of generic MEs.  We have shown that $K=3<K_{{\rm min}}=5$ PREs are ruled out for a number of $L>1,D=3$ MEs but the $K=4$ proof was beyond our current computing resources, and is a topic of future work.  Similarly, in the case of $D=3,L=1,K_{{\rm min}}=8$, we were only able to show that $K\leq 5$ was not generically possible, despite it being desirable to extend this to ruling out $K$ such that $K_{{\rm min}}>K>(D-1)^2+1=5$
%$K_{{\rm min}}\leq 6$ 
(in order to emphasize the inadequacy of the original heuristic of \crf{karasik2011many}).  These computational tasks involve obtaining Gr{\"o}bner bases (Nullstellensatz related), but the numerical search for PREs, via polynomial homotopy continuation methods, also becomes intractable for moderate $D$.  In this latter case, it is the shear number of potential solutions (the B{\'e}zout bound is exponential in $K_{{\rm min}}D^{2}\sim D^4$ and obtaining a tighter bound than this can be NP-hard) that proves difficult.  We recently become aware of the newly introduced monodromy extension~\cite{duff2016solving} to polynomial homotopy continuation that shows great promise for finding PREs.  Its advantage is that it quickly begins to track a non-optimal number of potential solutions, but with this non-optimality constrained to be linear in the actual number of solutions. 
One can then see that the complexity of finding a PRE would be linearly dependent upon the fraction of solutions to the system of polynomials that are real-valued and obey the positivity constraints for transition rates.  It is true that the difficulty of tracking each individual potential solution is still dependent upon $D$, but this scaling is not expected to be as severe as that of the bound on the number of solutions.  Despite its immediate implementation in this work, we feel that monodromy extension's capability is yet to be fully explored and we hope to do so in the future.  Another point, is that it may be possible to solve one generic system completely and then perform a `cheater's homotopy' to find PREs for other MEs, that have the same co-efficient structure, in a cheap fashion. 

As yet, there have been no experiments aiming to create PREs (involving non-orthogonal pure states). It certainly would be difficult to implement the high-efficiency adaptive measurement schemes required, 
though a suggestion has been made~\cite{daryanoosh2016stochastic}, involving quantum transport
with feedback, which alleviates some of the difficulties. However, PREs might have applications, irrespective of experimental realization, in quantum simulations. 
A generic quantum trajectory, which involves periods of non-trivial continuous evolution, will occupy an infinite sized ensemble of states, though this is reduced to a finite number when simulated using finite precision.  Despite this, the memory required will still be exponentially large in the system dimension, $D$.  In contrast, a finite PRE with only $K$ states, will have memory requirements scaling generically only as $D^2$ (see \erf{simpleKMin2}).  The catch, as the reader of this paper will appreciate, is that there is a one-time large resource cost associated with identifying a PRE (and adaptive measurement scheme) applicable to the ME.  This cost is dependent upon the applied algorithm, with the monodromy extension to polynomial homotopy continuation providing the greatest potential of allowing PREs to be found with sub-exponential (or a very small constant) complexity.  If the finding of PREs can be made tractable for the ME in question, then it is possible that trajectory simulation could be most efficiently undertaken using the PRE ME unraveling.

We have discussed the computational difficulty of our research and have suggested and applied some possible approaches to ameliorate this.  It is likely that in the future, more intensive or efficient studies can be undertaken. What then, are further topics that should be explored? Many of the characteristics of PREs are still unknown. In particular, PREs have been explicitly confirmed to exist only for very small systems.  Finding PREs for larger systems is perhaps the most fundamental task.  An interesting direction would be for cases (where there are more parameters than constraints)  in which there exists a positive dimensional PRE solution set (such as for $D=2$ where an extra transition was allowed, as discussed above). This freedom could be used to engineer some aspect of the PRE.  For example, perhaps one could specify values for some observables, or other properties of the ensemble members.  A study of the nature of the PREs themselves would also be of interest.  For example, how does the relative size of ME terms (e.g. Hamiltonian and Lindblad) influence the mean dwell time across the ensemble, as well as the ensemble entropy (both Shannon and Von Neuman)?
 
It is of arguably fundamental interest to know whether PREs exist in 100\% of cases where, according to our heuristic, there are more parameters than constraints. If this were true then there would be {\em only} a quadratic difference between the classical and (minimal) quantum ensemble size, despite the fact that there are infinitely many possible pure states for any finite $D$ that the system could explore. 
In fact, this would be the case even under the weaker condition that, for any $D$, there always exists a PRE for $K$ greater than $K_{{\rm min}}$ by a term $o(D^2)$ (it could be 1, but not necessarily so small). 
The quadratic difference, if it is universal, would presumably directly relate to the difference between a quantum density matrix, with 
$\sim D^2$ parameters, and a classical probability distribution, with $\sim D$ parameters, as arises from fundamental considerations of the difference between quantum and classical~\cite{quant-ph/0101012}.

It is perhaps appropriate to conclude this paper with a different perspective on the significance of there being a gap between $D$ and $K_{{\rm min}}$.  This gap makes open quantum systems harder to track than open classical systems, as per our title, but also suggests that there is a resource associated with an open quantum system.  Specifically, despite only having $D$ internal states, we have shown that generically, for $D>2$, a PRE can represent a finite classical hidden Markov model having $K>D$ states and therefore provide can a compression relative to the classical implementation~\cite{monrasPub}.  It seems likely that this compression could be arbitrarily large as increasing $K$ leads to a larger ratio of parameters to constraints.  Similarly, for a stochastic process with a fixed number of states that can be mapped to a PRE, a lower internal entropy implementation is possible as the PRE is comprised, in general, of non-orthogonal states~\cite{Gu2012}. Links between this work, and quantum machines more generally, will be explored elsewhere.  Also of interest would be to further explore the relation of PREs to aspects of emergent classical dynamics, including decoherence and chaotic systems. \blk

\acknowledgments
The authors acknowledge the HPC service at The University of Sydney for providing cluster computing resources that have contributed to the research results reported within this paper.  This work was supported by the Australian Research Council via discovery project number DP130103715, via the Centre of Excellence in Engineered Quantum Systems (EQuS), project number CE110001013 and CE170100009, and 
via the Centre for Quantum Computation and Communication Technology (CQC2T), project number CE110001027 and  CE170100012, and by the University of Sydney Faculty of Science via a Postgraduate Scholarship.  
Particular acknowledgment is given to the extremely generous assistance provided by both Jeff Sommars, regarding PHCpack and the monodromy method, and Dr Allan Steel with regards to the use of MAGMA.  We note that both Jeff and Allan provided custom scripts in order to more efficiently conduct our investigations. We thank Shuming Cheng for helpful comments on the manuscript. 
We acknowledge the traditional owners of the land on which this work was undertaken at Griffith University, the Yuggera people, and the traditional owners of the land on which this work was undertaken at the University of Sydney, the Gadigal people of the Eora Nation.

\appendix

\section{Incompletely connected graphs and Wigner symmetry}
\label{wignerGraphsApp}
\blk

In \srf{rateCounting} it was shown that linear algebra arguments can be used to rule out certain PREs; in particular highly connected graphs may not be consistent with dimensionality of the ensemble.  Such arguments have implications when searching for a PRE with a Wigner symmetry:
the graph representing the transitions of the potential PRE may not fully support the symmetry in question. 
In such cases, the asymmetric ensemble members would have to be mapped to themselves via $k^{\prime}=k$.  This leads to a less than maximal portion of the constraint equations being redundant.  This is perhaps best understood by example. 

Consider \frf{graphs}(d), where only the highly connected set of nodes $\{1,2,3\}$ would be expected to support a symmetry (node 3 is not equivalent to nodes $\{1,2,4\}$ as it receives a connection from node 6).  We then have (at least) 4 equivalence classes ($[\ket{\phi_1}]$, $[\ket{\phi_3}]$, $[\ket{\phi_5}]$ and $[\ket{\phi_6}]$) that we need to satisfy constraints for.  In this way, (at most) 2 values of $k$ can be chosen (from $\{1,2,4\}$) whose constraints are rendered redundant by the PRE symmetry.  This is described as `less than maximal' because some MEs could possess a symmetry that would allow fewer PRE equivalence classes, and hence more redundant equations. 
To illustrate this,
we consider a unitary Wigner symmetry, ${\cal T}$, which is defined by its action on any operator $\hat{O}\in\mathfrak{B}\left({ \mathbb{H}}\right)$ by ${\cal T}\hat{O}=\hat{U} \hat{O} \hat{U}^{\dag}$, with $\hat{U}$ the unitary operator $\ket{1}\bra{2}+\ket{2}\bra{3}+\ket{3}\bra{1}$.  This ME has an obvious cyclic permutation symmetry, $\mathbb{Z}_3$, and a fully connected graph of 6 nodes could support 2 equivalence classes, say $[\ket{\phi_1}]_{\sim}=\{\ket{\phi_1},\ket{\phi_2},\ket{\phi_3}\}$ and 
$[\ket{\phi_4}]_{\sim}=\{\ket{\phi_4},\ket{\phi_5},\ket{\phi_6}\}$, such that only 2 matrix constraint equations (1 from each class) need be considered.  This compares with the larger number ($4$) of classes required for the less symmetric graph of \frf{graphs}(d).  
The point being made is that there can be an incompatibility between the graph under consideration and the ME symmetry, which reduces the potential combined simplification of finding PREs. As such, the choice of graph should include consideration of available symmetry.

\section{Computational proofs of PRE non-existence}
\label{nullenstantz}
For a PRE to exist, it must be found as the solution to \erf{jumpCond} with the additional proviso that the transition rates, $\kappa_{jk}$, are real and positive.  As described in \srf{computationalDifficulty}, \erf{jumpCond} represents a system of polynomial constraints.  For the purposes of this appendix (which closely follows \cite{cox1992ideals}), let us label the parameters of \erf{jumpCond} as $\{x_{1},...,x_{n}\}$ and the set of constraining polynomials as $\{p_{1},...,p_{m}\}={\bf 0}$ (a vector of zeroes).  We next define the {\it affine variety}, $V(p_{1},...,p_{m})$, over the  complex field, $\mathbb{C}$, as the set of all solutions of the system of equations 
$\{p_{1},...,p_{m}\}={\bf 0}$.  If a PRE exists then it is necessary (but not sufficient) for there to be at least one solution $\in \mathbb{C}^{n}$ of \erf{jumpCond} and, resultantly, $V(p_{1},...,p_{m})\neq\emptyset$. Conversely, if $V(p_{1},...,p_{m})=\emptyset$ over the complex field then this is sufficient to rule out the possibility of PREs.  Note that $\mathbb{C}$ is an algebraically closed field.

Before giving the condition under which $V(p_{1},...,p_{m})=\emptyset$, we need to introduce the ${\it ideal}$ of $\{p_{1},...,p_{m}\}$.  This is defined by
\beq
I(p_{1},...,p_{m})=\langle p_{1},...,p_{m}\rangle =\{\sum_{i=1}^{s}h_{i}p_{i}: \ h_1,...,h_s\in\mathbb{C}[\vec{x}_n]\},
\eeq
where $s$ is any finite index and $\mathbb{C}[\vec{x}_n]$ is the set of all possible polynomials in variables $\{x_{1},...,x_{n}\}$ having complex coefficients.  From this definition it can be seen that if $1\in I$ then $I=\mathbb{C}[\vec{x}_n]\}$.  A crucial fact concerning ideals is that $I(p_{1},...,p_{m})$ shares the same zeroes as $\{p_{1},...,p_{m}\}$.  We now state the weak form of Hilbert's Nullstellensatz (which implies and is equivalent to the strong form): $I=\mathbb{C}[\vec{x}_n]$ iff $V(I)=\emptyset$, provided $\{x_{1},...,x_{n}\}$ belong to an algebraically closed field (in our case $\mathbb{C}$).  That is, if the variety of the ideal is empty then the ideal must be the set of all complex polynomials in variables $\{x_{1},...,x_{n}\}$.  However, we already know that this implies that $1\in I$.  Hence, showing that $1\in I$ is sufficient for excluding the possibility of PREs.  The task of showing that $1\in I$ is made easier by noting that for any monomial ordering $\{ 1\}$ is the only reduced Gr{\"o}bner basis for the ideal $\langle 1\rangle$~\cite{cox1992ideals}.  Fortunately, there are many sophisticated software packages that can calculate Gr{\"o}bner bases (we used MAGMA~\cite{bosma1997magma}).  This allowed us to computationally prove that no PREs exist for some specific MEs that are discussed in \srf{results}.  The reader wishing to learn the mathematical details concerning Gr{\"o}bner bases is referred to \cite{cox1992ideals}, while a discussion in the context of PREs is presented in the appendix of \crf{karasik2011tracking}.

\section{Polynomial homotopy continuation}
\label{PHCpackMonodromy}
It is likely that polynomial homotopy continuation (PHC) methods will find increasing use in the physical sciences going forward, especially since that they can easily be implemented in parallel (see~\cite{mehta2011numerical,mehta2011finding,mehta2016numerical} for some existing applications).  To understand how this arises we first pose the problem that we wish to solve: given $n$ polynomials ${\bf P}=\{p_{1},...,p_{n}\}$ in $n$ variables $\{x_{1},...,x_{n}\}$ (a square system) we want to find isolated solutions of $\{p_{1},...,p_{n}\}={\bf 0} $ (a vector of zeroes).  Although, in this paper, we were  satisfied with finding a single isolated solution, let us pose the question of finding all solutions (in any case the method will lead to an obvious algorithm to find a single one).  The homotopy continuation method of solving the polynomial system is to define a new polynomial system, ${\bf Q}=\{q_{1},...,q_{n}\}={\bf 0}$, in the same variables that is trivial to solve.  Then a family of polynomial systems, ${\bf H}$, is defined according to~\cite{li1997numerical,sommese2005numerical,verschelde1999algorithm}
\beq
{\bf H}=\gamma(1-t){\bf Q}+t{\bf P}={\bf 0},\quad\quad t\in[0,1]
\eeq
with $\gamma\in \mathbb{C}$.  It is clear that when $t=0$ the system ${\bf H}$ has the same solution set as the trivially soluble system ${\bf Q}$, while when $t=1$ it is equal to the target system of interest that we actually wish to solve.  It is natural to consider what happens to the solutions of ${\bf Q}$ as $t$ is moved from zero to one.  For ${\bf H}={\bf 0}$ to be a good homotopy we require that~\cite{li1997numerical}: 1) the solutions for $t=0$ be easy to find, 2) no singularities along the solution paths occur and 3) all isolated solutions can be reached.  Given these properties, the solution set of ${\bf H}$ can be tracked using standard techniques~\cite{allgower2012numerical} as $t$ is varied from zero to one.  Moreover, each of the paths can be tracked independently, leading to an extremely parallel implementation.  Remarkably, with proper care, it is possible to choose a homotopy such that the solution paths never cross, except perhaps a the end of the homotopy, $t=1$ (see\cite{sommese2005numerical} and ~\cite{verschelde1999algorithm} for a discussion of so called `end game' methods of dealing with these at $t=1$).  The inclusion of the complex valued $\gamma$ ensures that the paths are well behaved.  A crucial feature of ${\bf Q}$ required for the `good homotopy' conditions to be met is that it has as many or more isolated solutions as does ${\bf P}$.

The most naive way to choose the start system ${\bf Q}$, is based on the total degree, $D$, of ${\bf P}$.  If $d_{i}$ is the degree of the polynomial $p_{i}$ then the total degree is given by $D=\prod\limits_{i=1}^{n} d_{i}$.  B{\'e}zout's theorem states that ${\bf P}$ has at most $D$ isolated solutions in $\mathbb{C}^{n}$, so if ${\bf Q}$ is chosen in this manner all solutions are accessible.  The issue with this is that usually ${\bf P}$ has significantly fewer solutions than $D$, particularly if it is sparse in the sense that not all possible monomials appear in each polynomial in the set.  This is certainly the case for the constraints defined by \erf{jumpCond} as can be seen, for example, by noting that there are only linear powers of $\kappa_{jk}$.  The total degree based on \erf{jumpCond}, for $K=4$ in $D=3$ when the ME possesses a re3it invariant subspace, is about $5.6\times 10^{10}$ --- although much smaller than the completely generic ME, it is still too large for our purposes (for perspective, a system is defined as large for PHC purposes in~\crf{leykin2006parallel} as being when more than $5.6\times 10^{5}$ need to be tracked).  The key, therefore, is the establishment of a tighter bound than that of B{\'e}zout.

Unfortunately, the known tighter-than-B{\'e}zout bounds are themselves NP-hard to obtain~\crf{malajovich2007computing}.  
The direct consequence of this for our work was that when a black-box implementation of PHCpack~\cite{verschelde1999algorithm} was undertaken the computation did not return a start system in the allocated time for the $K=4$ size 24 polynomial system.
However, we made progress by investigation of the method of bounding that PHCpack used for smaller, but related, systems.  Specifically, the systems of~\cite{karasik2011many} were re-solved using PHCpack and it was noted that fairly tight bounds were being obtained from the $m$-homogenous B{\'e}zout number.  Without delving into the details (which can be found in the references already provided in this section), the important feature for our purposes is that this modified bound is obtained by partitioning the variables of the polynomial system into sets, with different bounds arising from different partitions.  Ideally, one would enumerate over all the sets, calculating each derived bound, but this is not possible due to there being $1.7\times 10^{9}$ of them with 24 variables.  Fortunately, the pattern from smaller $D=2$ systems was clear: a good partition is obtained by separating the variables representing the state vectors from the transitions rates.  Additionally, the transitions out of each node are grouped respectively.  This led to 5 partitions and, when evaluated, a $5$-homogenous B{\'e}zout number of $4.1\times 10^{7}$.  This is still huge, but is now tractable, particularly as we do not require the tracking all the potential solution paths.  We can stop once a solution meeting the physical criteria for a valid PRE is found {\it or} enough paths have been tracked that we believe the likelihood of a PRE existing for the example ME has been diminished to the extent that the search is aborted.  In the latter case, a new ME is chosen and the search restarted for the new system.

Due to the ad hoc implementation of the $m$-homogenous B{\'e}zout bound, with it being NP-hard to optimally obtain in general, it is desirable to obtain a methodology that avoids the necessity of its calculation.  This is achieved with the recently developed monodromy based solver described in \crf{duff2016solving}, the source code for which is available at \cite{monodromySource}.  The beauty of this method is two-fold in that solution paths can begun to be tracked after only a short pre-calculation {\it and} that the number of paths that need to be tracked is linear in the number of solutions of the target system~\cite{duff2016solving}.  This greatly increased the rate at which we could examine MEs for the existence of PREs.  Indeed it is perhaps possible to tackle larger systems than those for which we obtained PREs for in this paper.  Doing so represents an exciting future direction to explore.

\section{System details for reproducibility}
\label{repro}
In this section we ensure that our major claims are reproducible, and fully specified, by providing the parameter values for an example system from each class of ME appearing in \trf{systemSummary}.  Where appropriate, the parameters of an associated PRE are also given.  Although the task of finding a PRE is very difficult (NP-hard), verifying a provided solution is, by definition, only polynomially difficult and this is left for the reader to confirm.  We cannot, of course, provide a polynomially verifiable result when we obtain a Hilbert Nullstellensatz proof that a system has no PRE.  Instead we note that they can be achieved straightforwardly by calculation of the Gr{\"o}bner basis on computational algebra software~\cite{bosma1997magma}.  Although there were systems for which the computational proof took days, the examples that we have chosen to display are ones that only took minutes (or less).  To specify the $D=3$ MEs we provide the $\vec{\alpha},\vec{\gamma}^{l}$ which determine the Hamiltonian and Lindblad terms according to \erf{meParams}.  When a PRE has been found the ensemble states, $\{ \ket{\phi_k} \}_{k}$, transition rates, $\kappa$ (as a matrix) and the measurement scheme $\vec{\beta}^{k}$, ${\bm S}^{k}$ are provided.  This information is provided in \trf{systemSpec}.

\begin{table}[h!]
 \centering

\begin{tabular}{ccc}
 \centering
\textbf{Class}&\textbf{\specialcell[b]{ME, PRE,\\Meas. Scheme} } & \textbf{Values}\\ \hline

\multirow{7}{*}{I}& $\vec{\alpha}$& $\{0.06,  -0.47-0.47i,-0.04 -0.09i ,-0.48, 0.41-0.15i,-0.21\}$\\

& \multirow{2}{*}{$\vec{\gamma}^{1}$}

&$\{-0.54 - 1.94i, -2.71 + 2.07i, 2.93 + 1.01i,0.82 + 0.05i,$\\
&&$ 0.83 - 2.76i,-0.6 + 1.42i,0.44 - 1.71i, -1.47 - 0.42i\}$\\

& \multirow{2}{*}{$\vec{\gamma}^{2}$}

&$\{-2.73 - 2.68i, -1.71 - 1.26i, -2.85 + 1.27i,1.89 + 2.53i,$\\
&& $ 1.87 + 0.33i,-1.66 + 1.21i,-2.18 + 0.72i, -0.11 + 0.53i\}$\\

& \multirow{2}{*}{$\vec{\gamma}^{3}$}

& $\{-1.18 - 2.69i, 2.01 + 1.03i, 2.73 - 2.71i,-1.11 + 1.35i,$\\
&&$ -0.11 - 0.67i, -1.41 - 2.12i,-0.64 + 0.15i, 0.82 - 0.88i\}$\\

\hline
\multirow{3}{*}{II}&${\vec{\alpha}}$& $\{-1.24, - 0.79+1.17i,0.06 -0.09i,1.39,0.09  -1.15-0.09i,1.1\}$\\

& \multirow{2}{*}{$\vec{\gamma}^{1}$}

&$\{2.06 - 0.95i, -2.55 - 0.1i, 0.81 - 0.11i,-1.32 - 2.35i,$\\
&&$-1.33 - 2.9i, 2.1 + 2.11i,1.02 - 2.99i, -2.72 - 2.74i\}$\\

\hline

\multirow{14}{*}{III}&${\vec{\alpha}}$& $\{0,- 0.81i,- 2.2i,0,- 0.39i,0\}$\\
&${\vec{\gamma}^{1}}$& $\{1.22, 0.48, 2.67,-2.1, -1.04, 2.3,-0.01, 0.33\}$\\
&${\vec{\gamma}^{2}}$& $\{2.75, 1.93, -1.93,2.35, -0.25, -2.87,2.46, 0.93\}$\\
&${\vec{\gamma}^{3}}$& $\{-2.95, -0.24, 0.08,-1.15, -2.67, -2.3,2.07, 0.94\}$\\

& \multirow{2}{*}{$\{ \ket{\phi_k} \}_{k}$}

& $\{\{0.371, -0.884, 0.283\}, \{0.046, 0.999, -0.011\},$\\
&&$\{0.338, -0.795, -0.504\}, \{-0.651, 0.756, -0.069\}\}$\\

%&$\{ \ket{\phi_k} \}_{k}$& $\{\{0.676, -0.372, -0.637\}, \{-0.423, -0.703, -0.571\},\{-0.423, 0.703, 
%-0.571\}, \{0.676, 0.372, -0.637\}\}$\\

& \multirow{2}{*}{\mbox{\boldmath$\kappa$}}

& $\{\{0, 21.777, 16.12, 9.521\},\{4.809, 0, 17.162, 1.057\}, $\\
&&$\{5.513, 1.883, 0, 5.155\}, \{13.97, 1.599, 12.946, 0\}\}$\\

%&\mbox{$\kappa$}& $\{\{0, 0.002, 0.066, 0.062\},\{0.067, 0, 0.005, 0.041\}, \{0.041, 0.005, 
%0, 0.041\}, \{0.062, 0.066, 0.002, 0\}\}$\\

& \multirow{2}{*}{$\{\vec{\beta}^{k} \}_{k}$}

& $\{\{1.143, 1.52, -2.256\},\{-2.743, -3.073, -2.626\}, $\\
&&$\{-0.509, -3.317, -2.43\},\{0.71, -0.005, 4.22\}\}$\\

%&$\{\vec{\beta}^{k} \}_{k}$& $\{\{0.004, 0.189, 0.128\},\{0.122, 0.199, 0.127\},\{-0.199, 0.127, 0.122\},\{0.189, 0.004, -0.128\}\}$\\

&\mbox{\boldmath $S$\unboldmath $^{1}$}& $\{\{-0.925, 0.093, -0.369\}, \{0.355, 0.56, -0.748\}, \{-0.137, 0.823, 0.552\}\}$\\
&\mbox{\boldmath $S$\unboldmath $^{2}$}& $\{\{0.52, 0.814, 0.257\},\{0.41, 0.026, -0.912\}, \{-0.749, 0.58, -0.32\}\}$\\
&\mbox{\boldmath $S$\unboldmath $^{3}$}& $\{\{-0.375, 0.784, 0.495\}, \{-0.348, -0.614, 0.708\}, \{0.859, 0.093, 0.503\}\}$\\
&\mbox{\boldmath $S$\unboldmath $^{4}$}& $\{\{0.722, -0.483, 0.495\}, \{-0.135, -0.8, -0.584\}, \{-0.678, -0.355, 0.643\}\}$\\

\hline

\multirow{13}{*}{IV}&${\vec{\alpha}}$& $\{0,0,-0.04,0,0,0\}$\\
&${\vec{\gamma}^{1}}$& $\{0,-0.38,0,0,0,-0.38,0,0\}$\\
&${\vec{\gamma}^{2}}$& $\{0,0,0,0,0,0,0.43,0\}$\\

& \multirow{2}{*}{$\{ \ket{\phi_k} \}_{k}$}

& $\{\{0.676, -0.372, -0.637\}, \{-0.423, -0.703, -0.571\},$\\
&&$\{-0.423, 0.703, -0.571\}, \{0.676, 0.372, -0.637\}\}$\\

%&$\{ \ket{\phi_k} \}_{k}$& $\{\{0.676, -0.372, -0.637\}, \{-0.423, -0.703, -0.571\},\{-0.423, 0.703, 
%-0.571\}, \{0.676, 0.372, -0.637\}\}$\\

& \multirow{2}{*}{\mbox{\boldmath$\kappa$}}

& $\{\{0, 0.002, 0.066, 0.062\},\{0.067, 0, 0.005, 0.041\}, $\\
&&$\{0.041, 0.005, 0, 0.041\}, \{0.062, 0.066, 0.002, 0\}\}$\\

%&\mbox{$\kappa$}& $\{\{0, 0.002, 0.066, 0.062\},\{0.067, 0, 0.005, 0.041\}, \{0.041, 0.005, 
%0, 0.041\}, \{0.062, 0.066, 0.002, 0\}\}$\\

& \multirow{2}{*}{$\{\vec{\beta}^{k} \}_{k}$}

& $\{\{0.004, 0.189, 0.128\},\{0.122, 0.199, 0.127\}, $\\
&&$\{-0.199, 0.127, 0.122\},\{0.189, 0.004, -0.128\}\}$\\

%&$\{\vec{\beta}^{k} \}_{k}$& $\{\{0.004, 0.189, 0.128\},\{0.122, 0.199, 0.127\},\{-0.199, 0.127, 0.122\},\{0.189, 0.004, -0.128\}\}$\\

&\mbox{\boldmath $S$\unboldmath $^{1}$}& $\{\{0.76, 0.519\}, \{-0.298, 0.813\}, \{0.577, -0.264\}\}$\\
&\mbox{\boldmath $S$\unboldmath $^{2}$}& $\{\{0.313, -0.216\},\{0.424, -0.839\}, \{0.85, 0.499\}\}$\\
&\mbox{\boldmath $S$\unboldmath $^{3}$}& $\{\{0.424, 0.839\}, \{-0.85, 0.499\}, \{-0.313, -0.216\}\}$\\
&\mbox{\boldmath $S$\unboldmath $^{4}$}& $\{\{0.298, 0.813\}, \{-0.76, 0.519\}, \{0.577, 0.264\}\}$\\

\end{tabular}
\caption{An example system for each class of \trf{systemSummary} is specified. The ME is determined by $\vec{\alpha},\vec{\gamma}^{l}$, while PREs (if they exist) are given in terms of the ensemble states, $\{ \ket{\phi_k} \}_{k}=\{ \ket{\phi_1},\ket{\phi_2},\ket{\phi_3},\ket{\phi_4} \}$ and the transition rates \mbox{\boldmath$\kappa$} $\equiv\kappa_{jk}$ (with the diagonal elements irrelevant and set to zero).  We represent the re3it as $\{x,y,z\}$ with $\ket{\phi}=x\ket{1}+y\ket{2}+z\ket{3}$.  For each PRE, an explicit adaptive measurement scheme that can realize it is provided via the parameters $\vec{\beta}^{k}$, ${\bm S}^{k}$.  The ME parameters are exact but the PRE and measurement scheme are necessarily approximations that can be refined to arbitrary accuracy if desired.  Both class III and IV possess the re3it invariant subspace symmetry as the Lindbladians are real-valued.  This leads to a real-valued PRE and measurement scheme.  Class IV additionally possesses the Wigner unitary symmetry $\hat{U}=\mathbbm{1}-2\ket{2} \bra{2}$, which is reflected in the symmetry of the PRE, as per \erfs{induced}{inducedMap}, and the measurement scheme.
}
\label{systemSpec}
\end{table}

%\newpage

\section{Finding measurement schemes that realise a PRE}
\label{finMeasScheme}
Given a viable PRE [an ensemble $\{ \kappa_{jk}\geq 0,\, \ket{\phi_k} \}$ satisfying \erf{jumpCond}], it is expeceted that there exists an appropriately applied measurement scheme such that the conditioned state of the quantum system will, in the long-time limit, jump between the ensemble members, spending a time in the state $\ket{\phi_k}$ proportional to $\wp_k$.  The measurement scheme in question will, in general, be adaptive, meaning that the experimental setting parameters ($\vec{\beta}$, ${\bm S}$), must be changed according to which state 
($k$) the system is currently in.  To find the explicit measurement scheme parameterization we construct the evolution for the PRE.  The pre-jump state $\ket{\phi_k}$ will be transformed by a `click' in the $m$th detector into one of the states, labeled by the integer $f(m,k)$, comprising the ensemble.  Note that $f(m,k)$ could be equal for different $m$ and has a range that includes $k$ (as in $f(m,k)=k$ is possible).  The rate $\kappa_{jk}$ is, therefore, interpreted as the total transition rate from state $k$ to state $j$ arising from possibly multiple detection channels, which can be made explicit via $\kappa_{jk}=\sum_{m=1}^{M}\delta_{j,f(m,k)}\lambda^{k}_{m}$.  The rates, $\lambda^{k}_{m}$, in this expression reference the transition rate out of state $k$ due to detector $m$.  Thus, we describe the post-jump state, resulting from an $m$th detector click given the pre-jump state $\ket{\phi_k}$, as
\beq 
 \hat{c}_{m}^{\prime} (k)\ket{\phi_k}\equiv \hat{c}_{m}^{k}\ket{\phi_k}=\left(\sum_{l=1}^{L} S_{ml}^{k} \hat{c}_l + \beta^{k}_m\right) \ket{\phi_k} \propto \ket{\phi_{f(m,k)}},
\label{cycliciff}
\eeq
while the no-jump evolution proceeds according to 
\beq
\hat{H}^{k}_{\rm eff} \ket{\phi_k}\propto\ket{\phi_{k}}.
\label{wings}
\eeq  We have introduced some extra notation to make it explicit that an adaptively modified measurement setting $\vec{\beta}^{k}$, ${\bm S}^{k}$ exists for each state $ \ket{\phi_k}$ in the ensemble.  As an aside, note that the measurement parameters for different $k$ can be directly related in the case that the measurement scheme possesses Wigner symmetry (see \cite{warwis2018a} for further details).

Let us now discuss the proposed method of solving \erfa{cycliciff}{wings} to find an explicit measurement scheme that realizes a given PRE.  The most important thing to note is that the PRE, in particular $\{\ket{\phi_k}\}$,  is assumed to have already been found, via solution of \erf{jumpCond}.  This leaves the following as variables: $\vec{\beta}^{k}$, ${\bm S}^{k}$ and $f(m,k)$, for all $k$ (and all $m$ in the last case).  Note that the proportionality constants can be determined from the, assumed also to have been already found, transition rates, $\kappa_{jk}$.  The alternative approach, of solving \erfa{cycliciff}{wings} simultaneously for the PRE and measurement scheme (making unnecessary the solution of \erf{jumpCond}), is likely to be much harder, in general, due to the highly non-linear scaling of the difficulty of solving systems of polynomials in the number of variables.  The explicit measurement schemes that we find in this paper have the property that $f(m,k)\neq f(m^{\prime},k)\neq k$ for all $m\neq m^{\prime}$, so some of the generic difficulty of solving \erfa{cycliciff}{wings} is avoided.  A more exhaustive examination of the range of possible measurement schemes that can achieve a particular PRE is of interest for future work.

As a final technical point, because the specification of the PRE involves machine precision numbers (that is, they are not exact), it will typically be the case that \erfs{cycliciff}{wings} have no solution (due to rounding).  Instead, the constraints can be formulated as a minimisation problem.  The approximate solution can be made arbitrarily accurate by increasing the precision of the numerically defined PRE.

\bibliographystyle{unsrtnat}

\bibliography{bibliography9HRef}

\begin{thebibliography}{51}
\bibitem[sch()]{schemeComment}
It is important to note that the implied number of detectors (equal to $M$) can
  be greater than the number of Lindblad operators, $L$, describing the ME.
  This is most easily understood in a quantum optics context: linear
  interferometers, described by ${\bm S}$, take the field outputs of the system
  as inputs while $\vec{\beta}$ represents the addition of weak local
  oscillators (WLOs) to the interferometer outputs prior to photodetection. The
  merging and splitting of system output fields can also refer to frequency
  conversion, so that the `linear interferometer' is to be understood in the
  most general sense. The LOs are required to be weak (that is, $|{\beta}_l|^2$
  not much larger than $\Tr[c\dg_l c_l \rho_{\rm ss}]$) \blk so that a
  non-negligible amount of information is gathered concerning the system upon
  each detection, meaning that the quantum trajectory is jump-like rather than
  diffusive; consequently, the PRE can consist of a finite number of states. We
  stress that most jump-like unravellings will lead, like diffusive
  unravellings, to PREs of infinite size. Those that lead to finite PREs, as we
  are interested in, are exceptional.
\bibitem[ran()]{rankMEcomment}
In the case that the rank of $\rho_{{\rm ss}}$ is less than $D$, then we can
  project the Lindblad ME onto a smaller dimensional subspace and redefine that
  as D. Also note that uniqueness of $\rho_{{\rm ss}}$ is assumed in our work
  for convenience. If the assumption were not made then our results would
  become dependent upon the particular steady-state under consideration. For
  example, different steady-states could be of differing rank, and also have a
  differing effective number of decoherence channels (non-zero
  $\Tr[\hat{c}\dg_l\hat{c}_l \rho_{{\rm ss}}]$). These complications could be
  included, but would distract from the underlying physics that we wish to
  highlight.
\bibitem[tra()]{trackingComment}
Note that this `tracking', a numerical technique of polynomial homotopy
  continuation, is unrelated to the efficient `tracking' of open quantum
  systems, which is the physics problem under investigation.
\bibitem[sub()]{subOptimalRates}
To prove that $L=1$ PREs are not possible for $K=4,5$ for the example MEs
  chosen, we rely upon ruling out various graphs using the considerations of
  \srf{rateCounting}. For the graphs that are not ruled out in this way, we
  then obtain computational Nullstellens{\"a}tze showing that no PRE can exist.
  In contrast to the case where a Nullstellensatz is obtained for a fully
  connected graph (for the $K=3$), for these $L=1$, $K=4,5$ examples there are
  many isomorphically different graphs that can be drawn when the number of
  non-zero rates is less than the maximal number. That is, ruling out the graph
  with the optimal (largest) number of transitions is not sufficient to prove
  that no PRE can exist. Instead, we must consider all the isomorphically
  different graphs with fewer rates also and then either rule them out via
  \srf{rateCounting} or by Nullstellensatz. This difficulty does not arise if a
  Nullstellensatz can be performed on the fully connected graph as all graphs
  with fewer transitions can be obtained from this graph just by setting some
  of the transition rates to zero. Consequently, one must first identify all
  the relevant graphs (there are about 50) and then obtain Nullstellens{\"a}tze
  for them in order to have an overall proof of $K\leq 5$ PRE non-existence for
  a single example ME. Some graph Nullstellens{\"a}tze take longer than others,
  with a total runtime for each ME of around 2 days.
\end{thebibliography}

\end{document}